\documentclass[aps,prd,groupedaddress,onecolumn]{revtex4-2}
\usepackage[utf8]{inputenc}

\usepackage{amssymb,amsmath,amsfonts}
\usepackage{mathtools}
\usepackage{mathrsfs}
\usepackage{bbm}
\usepackage{slashed}
\usepackage{nicefrac}

\usepackage{graphicx}
\usepackage[dvipsnames]{xcolor}
\usepackage{array}

\usepackage{simplewick}

\usepackage{hyperref}
\usepackage{xparse}
\usepackage{xspace}

\usepackage{tikz}
\usepackage{empheq}

\usetikzlibrary{decorations.pathmorphing}
\usetikzlibrary{automata,positioning}

\usepackage{cancel}
\usepackage[normalem]{ulem}

\usepackage{xifthen}
\usepackage{dsfont}
\usepackage[titletoc]{appendix}
\usepackage{booktabs}
\usepackage{units}

\newcommand{\gettitle}{}
\hypersetup{linkcolor=black
	colorlinks,
	linkcolor={red!75!black},
	citecolor={blue!75!black},
	urlcolor={blue!75!black},
	pdftitle={\gettitle},
	pdfauthor={Mehtar-Tani},
	pdfkeywords={Perturbative QCD} {Small-x},
	bookmarksopen=true,
	bookmarksopenlevel=2,
	bookmarksnumbered=true
}

\def\del{\partial}

\newcommand{\eqn}[1]{Eq.~\eqref{#1}}

\newcommand{\secn}[1]{Sec.~\ref{#1}}

\newcommand{\nn}{\nonumber\\ }

\def\be{\begin{eqnarray*}}
\def\ee{\end{eqnarray*}}
\def\beq{\begin{eqnarray}}
\def\eeq{\end{eqnarray}}

\newcommand{\bea}{\beq \begin{aligned}}
\newcommand{\eea}{\end{aligned}\eeq}

\newcommand{\bxi}{{\boldsymbol x_1}}
\newcommand{\bxii}{{\boldsymbol x_2}}
\newcommand{\bxiii}{{\boldsymbol x_3}}

\newcommand{\br}{{\boldsymbol r}}
\newcommand{\bx}{{\boldsymbol x}}

\newcommand{\bz}{{\boldsymbol z}}

\newcommand{\bk}{{\boldsymbol k}}
\newcommand{\bq}{{\boldsymbol q}}

\newcommand{\cP}{{\cal P}}
\newcommand{\cO}{{\cal O}}

\newcommand{\cM}{{\cal M}}
\newcommand{\cU}{{\cal U}}
\newcommand{\fscal}{\mu}
\newcommand{\rme}{{\rm e}}
\newcommand{\rmd}{{\rm d}}

\newcommand{\rmTr}{{\rm Tr}}
\newcommand{\rmtr}{{\rm tr}}

\def\abar{\bar\alpha_s}

\begin{document}

%\preprint{}


\title{Collinear Structure of Nonlinear Small-$x$ Evolution}


\author{Renaud Boussarie}
\email{renaud.boussarie@polytechnique.edu}
\affiliation{CPHT, CNRS, \'Ecole polytechnique, Institut Polytechnique de Paris, 91128 Palaiseau, France}
\author{Paul Caucal}
\email{caucal@subatech.in2p3.fr}
\affiliation{SUBATECH UMR 6457 (IMT Atlantique, Universit\'e de Nantes, IN2P3/CNRS), 4 rue Alfred Kastler, 44307 Nantes, France}
\author{Yacine Mehtar-Tani}
\email{mehtartani@bnl.gov}
\affiliation{Physics Department, Brookhaven National Laboratory, Upton, NY 11973, USA}


\date{\today}

\begin{abstract}

We introduce a novel approach to high-energy QCD factorization of cross-sections for processes involving a dilute projectile and a dense target. Our method preserves the factorization between ``fast" and ``slow" modes in the longitudinal momentum $k^+$ for a projectile moving along the positive light-cone direction, making it compatible with higher-order loop computations done within the standard formulation of the Color Glass Condensate (CGC) effective field theory. Moreover, it eliminates the anomalous double collinear logarithms that typically hinder the convergence of perturbation theory at small-$x$.
Our scheme amounts to a change of basis in the space of CGC operators, introducing an arbitrary transverse scale dependence while leaving physical cross-sections invariant. Implementing this scheme requires a modification of the high-energy renormalization group equation for the CGC operators, which we explicitly derive at next-to-leading order (NLO)
and in the non-linear regime for the case of the dipole operator and the DIS impact factor.
This general framework provides a consistent and systematic prescription for computing observables in the small-$x$ regime of QCD, free from collinear instabilities.
\end{abstract}

\keywords{Perturbative QCD, factorization, small-x, resummation, BK, BFKL, gluon saturation, CGC}


\maketitle

\section{Introduction\label{sec:intro}}

Quantum Chromodynamics (QCD) factorization provides the foundational framework for studying hadronic structure in high-energy processes \cite{Collins:1989gx}. A paradigmatic example is Deep Inelastic Scattering (DIS), characterized by a large momentum transfer $Q^2$ and a fixed Bjorken variable $x_{\rm Bj} =Q^2/s$, where $s$ is of the order of the squared center-of-mass energy of the virtual photon-nucleon collision. In this kinematic regime, the leading power in $Q^2$ factorizes into the convolution of a hard matrix element --- calculable in perturbative QCD --- that encodes short-distance dynamics involving quark and gluon fields, with universal parton distribution functions (PDFs), which capture the long-distance, non-perturbative structure of the hadron. Higher precision is systematically achieved by incorporating corrections at higher orders in the strong coupling $\alpha_s$.

This factorization picture has been remarkably successful in probing hadron structure. However, it is expected to break down at small $x_{\rm Bj}$ (or equivalently, in the high-energy or Regge limit $s \gg Q^2$~\cite{Regge:1959mz}), where large logarithms of $\ln(1/x_{\rm Bj})$ spoil the convergence of the perturbative series. Moreover, nonlinear effects become significant due to the rapid rise of gluon densities at small $x_{\rm Bj}$, signaling the onset of gluon saturation \cite{Mueller:1985wy,Gribov:1984tu,McLerran:1993ni}.

To address this breakdown, alternative frameworks have been developed, notably $k_t$-factorization~\cite{Catani:1990eg} and the dipole formalism in transverse coordinate space~\cite{Mueller:1993rr} (see e.g.~\cite{Gelis:2010nm} for a review). In these approaches, the central object is the unintegrated gluon distribution or, equivalently, the dipole $S$-matrix. Its evolution is governed by the Balitsky–Fadin–Kuraev–Lipatov (BFKL) equation in the dilute regime~\cite{Lipatov:1976zz,Kuraev:1977fs,Balitsky:1978ic} and by the nonlinear Balitsky-Kovchegov and Jalilian-Marian, Iancu, McLerran, Weigert, Leonidov and Kovner (BK–JIMWLK) equations in the dense regime~\cite{Balitsky:1995ub,Kovchegov:1999yj,JalilianMarian:1997jx,JalilianMarian:1997gr,Kovner:2000pt,Iancu:2000hn,Iancu:2001ad,Ferreiro:2001qy}. These evolution equations incorporate saturation physics~\cite{Gribov:1984tu,Mueller:1985wy,McLerran:1993ni}, a manifestation of unitarity at small $x$, and form the foundation of many phenomenological studies at HERA, RHIC, and the LHC (see e.g.~\cite{Morreale:2021pnn} for a review).

A major milestone in this theoretical development was the derivation of the next-to-leading order (NLO) corrections to both the BK equation~\cite{Balitsky:2007feb,Balitsky:2009xg} and the DIS impact factor~\cite{Balitsky:2010ze,Beuf:2011xd,Beuf:2016wdz,Beuf:2017bpd,Hanninen:2017ddy}, aimed at improving the precision of small-$x$ predictions. However, two related issues soon emerged, raising concerns about the consistency and reliability of the high-energy factorization framework: (i) the NLO BK equation was found to suffer from severe numerical instabilities~\cite{Lappi:2015fma}, (ii) the NLO impact factor for the total DIS cross-section defined as the finite part after naively subtracting the projectile rapidity divergence, can become large~\cite{Beuf:2014uia} leading to unphysical results such as negative cross-sections (see also~\cite{Stasto:2013cha,Ducloue:2016shw} for a similar phenomenon in forward particle production in 
proton-nucleus collisions, as well as the subsequent literature addressing this issue~\cite{Altinoluk:2014eka, Kang:2014lha, Stasto:2014sea, Watanabe:2015tja, Stasto:2016wrf, Iancu:2016vyg, Ducloue:2017mpb, Ducloue:2017dit, Xiao:2018zxf, Liu:2019iml, Liu:2020mpy,Shi:2021hwx}). For less inclusive processes, such as two-particle correlations in DIS, similar problems have been identified~\cite{Taels:2022tza,Caucal:2022ulg,Caucal:2023fsf} in connection with the interplay between small-$x$ and Sudakov resummation~\cite{Mueller:2012uf,Mueller:2013wwa}.

These instabilities have antecedents in the context of NLO BFKL evolution and are now understood to arise from an improper identification of rapidity logarithms $\ln(s/s_0)$, which require all-order resummation~\cite{Andersson:1995ju,Kwiecinski:1996td,Kwiecinski:1997ee,Salam:1998tj,Ciafaloni:1998iv,Ciafaloni:1999yw,Ciafaloni:2003rd,SabioVera:2005tiv,Motyka:2009gi}. At NLO, this manifests in the appearance of large collinear double logarithms of the form $\ln^2(Q^2/Q_0^2)$, which compromise the convergence of the perturbative expansion, particularly when the energy scale is chosen as $s_0 = Q_0^2$ or $s_0=Q_0Q$ with $Q_0 \ll Q$ a characteristic non-perturbative scale of the target. In the context of DIS for instance, identifying the rapidity logarithm with the choice $s_0=Q_0Q$ \footnote{The choice $s_0=Q_0Q$ is more natural in symmetric processes such as $\gamma^\ast \gamma^\ast$ scattering.} and relating it to $x_{\rm Bj}=Q^2/s$ yields 
\begin{equation}
\ln\left(\frac{s}{Q_0Q}\right) = \ln\left(\frac{1}{x_{\rm Bj}}\right) +\frac{1}{2} \ln\left(\frac{Q^2}{Q_0^2}\right),
\end{equation}
highlighting the fact that a change in the rapidity variable induces spurious collinear logarithms that become problematic beyond leading order when $Q\gg Q_0$. This feature can be traced back to a violation of time ordering in gluon emissions that is not systematically accounted for in the standard high energy factorization framework \cite{Beuf:2014uia,Ducloue:2019ezk}. This underscores the need for a consistent rapidity-factorization scheme that minimizes the sensitivity of the perturbative series on the choice of rapidity variable.
In the BFKL framework, such discrepancies are typically resolved by adjusting the energy scale $s_0$. However, in the nonlinear BK case, the resolution is more subtle, as no fully analytic treatment is available.

In DIS, the transverse scale of the photon impact factor is $Q^2 \gg \Lambda_{\rm QCD}^2$, or $\gg$ the saturation scale $Q_s^2$, implying a large phase space for strongly ordered gluon emissions:
\begin{equation}
Q_s \sim Q_0 \ll k_{\perp,1} \ll k_{\perp,2} \ll \cdots \ll Q\,.
\end{equation}
This DGLAP-like regime breaks the symmetry between $Q$ and $Q_0$ underlying the Regge factorization, which deals solely with $\ln s$ resummation, and thus, requires the resummation of contributions of the form $(\bar{\alpha}_s \ln(Q^2/Q_0^2))^n$ to all orders in the coupling constant $
\bar \alpha_s \equiv \alpha_s N_c/\pi$. In DIS, a more judicious choice for the rapidity variable is obtained by setting $s_0 = Q^2$, which ensures a proper resummation of the large collinear logarithms. In terms of rapidity ordering, this corresponds to evolution in $k^-$ rather than $k^+$~\cite{Ducloue:2019ezk}. In processes like $\gamma^* \gamma^*$ scattering, both collinear ($Q \gg Q_0$) and anti-collinear ($Q \ll Q_0$) regimes must be treated symmetrically, motivating the scale choice $s_0 = QQ_0$. Nevertheless, the shockwave formalism commonly adopts $k^+$ as the rapidity variable, which resums anti-collinear logarithms but not collinear ones, making it suboptimal for DIS.

To illustrate this, consider the one-loop correction to DIS in the dipole frame~\cite{Kopeliovich:1981pz, Bertsch:1981py, Mueller:1989st,Nikolaev:1990ja}, where the virtual photon propagates in the $+z$ direction. The natural evolution variable in this frame is the longitudinal momentum $k^+ \ll q^+$, where $q^+$ is the large light-cone component of the photon momentum. At double logarithmic accuracy, the phase space integral for gluon radiation reads:
\begin{equation} \label{eq:dl-ps}
\bar{\alpha}_s \int_{Q_0^2}^{Q^2} \frac{d k_\perp^2}{k_\perp^2} \int_{q_0^+}^{(k_\perp^2/Q^2)q^+} \frac{dk^+}{k^+} 
= \bar{\alpha}_s \ln\left(\frac{Q^2}{Q_0^2}\right) \ln\left(\frac{s}{Q_0Q}\right)=\bar\alpha_s\ln\left(\frac{Q^2}{Q_0^2}\right)\ln\left(\frac{1}{x_{\rm Bj}}\right)-\frac{\bar\alpha_s}{2}\ln^2\left(\frac{Q^2}{Q_0^2}\right),
\end{equation}
where $q_0^+$ is a target scale playing the role of a plus longitudinal momentum cut-off, and which can be taken as $Q_0^2/(2P^-)$, while $s=2q^+P^-$ is related the virtual photon-nucleon center-of-mass energy squared $W^2$ and the nucleon mass $m$ by $W^2=s-Q^2+m^2$. Throughout, we implicitly work in the Regge kinematics, where one assumes that $W^2\sim s$. 

The first term in \eqn{eq:dl-ps} is expected in the collinear limit from DGLAP resummation. However, the second collinear double logarithm --- arising from a mistreatment of the collinear limit in high-energy factorization --- is incorrect, and in fact incompatible with the renormalization group equations, which constrain the highest power of collinear logarithms to not exceed that of the coupling constant  \cite{Salam:1998tj}. 

The $k^+$ integral in Eq.\,\eqref{eq:dl-ps} is bounded from above by $(k_\perp^2/Q^2)q^+ $ as consequence of the gluon lifetime $t_f\sim 1/k^-$ being smaller than that of the virtual photon $1/|q^-|$, a feature that is correctly taken into account by NLO calculations in the CGC framework. However, CGC calculations in dilute-dense collisions typically neglect that $k^-$ is also bounded from above by the target longitudinal momentum $P^-$. This is a consequence of taking the shock-wave approximation where the longitudinal size of the target $\sim 1/P^-$ is taken to be zero, such that the constraint $k^-\sim k^+/k_\perp^2\ll P^-$ becomes irrelevant.

The kinematic constraint on $k^+$, 
\begin{equation} \label{eq:k-ordering}
k^+\gg \frac{k_\perp^2}{2P^-}\,,
\end{equation}
ensures proper time ordering between the gluon and the longitudinal size of the target. Enforcing this constraint modifies the phase space to:
\begin{equation} \label{eq:dl-ps-2}
\bar{\alpha}_s \int_{Q_0^2}^{Q^2} \frac{d k_\perp^2}{k_\perp^2} \int_{k_\perp^2 / 2P^-}^{k_\perp^2q^+/Q^2} \frac{dk^+}{k^+} 
= \bar{\alpha}_s \ln\left(\frac{Q^2}{Q_0^2}\right) \ln\left(\frac{s}{Q^2}\right)=\bar\alpha_s\ln\left(\frac{Q^2}{Q_0^2}\right)\ln\left(\frac{1}{x_{\rm Bj}}\right)\,,
\end{equation}
which differs from Eq.~\eqref{eq:dl-ps} by the argument of the rapidity logarithm. As compared to Eq.~\eqref{eq:dl-ps}, Eq.~\eqref{eq:dl-ps-2} has the expected double logarithmic structure, without additional collinear double logarithmic.

In order to resum both collinear and anti-collinear logarithms simultaneously in the BK equation, various strategies have been devised, including explicit resummation of leading double (and single) collinear logarithms and kinematically constrained evolutions~\cite{Beuf:2014uia,Iancu:2015vea,Ducloue:2019ezk}, which enforce ordering in both $k^+$ and $k^-$. 

While these methods alleviate numerical instabilities~\cite{Ducloue:2019ezk,Lappi:2016fmu}, they involve \textit{ad hoc}, albeit physically motivated, modifications and do not provide a systematic prescription for higher-order calculations.

In this work, we propose a robust framework for high-energy factorization based on evolution in the commonly adopted rapidity variable $\ln(k^+/q^+)$. Our approach exploits the factorization-scheme invariance of the cross section to redefine the dipole operator, its rapidity-evolution kernel, and the impact factor. This transformation introduces a new factorization scale $\mu$ that interpolates between collinear and rapidity logarithms, thereby reshuffling them among different components of the factorized cross section. Remarkably, the choice $\mu = Q$ matches the definition of the rapidity variable that restores proper time ordering at NLO and provides a consistent path toward stable higher-order evolution. This transformation ensures that collinear double logarithms are resummed to all orders. The resulting anti-collinear double logarithms, while present, are effectively neutralized by saturation dynamics, and thus do not compromise the stability of the evolution~\cite{Ducloue:2019ezk}. 

Our paper is organized as follows. Sec.~\ref{sec:results} summarizes the main results of this work; it is designed to be self-contained, compiling all practical formulas to be used for the numerical implementation of our high-energy factorization framework in the specific case of the DIS structure functions (the generalization to other standard small-$x$ processes depending on the dipole operator should be straightforward). Sec.~\ref{sec:he-factorization} motivates the introduction of our composite dipole formulation of the high-energy Operator Product Expansion (OPE) through an explicit computation of the one-loop correction to a generalized dipole operator, first introduced in~\cite{Boussarie:2021wkn,Boussarie:2020fpb}. This operator reduces to the standard dipole operator in the $x \to 0$ limit but also captures information about the longitudinal size of the shock wave, and hence about life-time ordering. In Sec.~\ref{sec:comp-dip}, we provide a formal treatment of the high-energy OPE in terms of this composite dipole, highlighting the structure of high-energy factorization and how it transforms under a change of basis between ``bare" and ``composite" operators. Secs.~\ref{sec:HNLO} and~\ref{sec:mellin-space} respectively address how this new high-energy OPE resolves issues related to small-$x$ impact factors and instabilities in the NLO renormalization group equation associated with collinear logarithms. Finally, in Sec.~\ref{sec:sudakov}, we derive the NLO BK equation for the composite dipole in coordinate space, which constitutes the essential practical component of high-energy QCD factorization in the non-linear regime.

%%%%%%%%%%%%%%%%%%%%%%%%%%%%%%%%%%%%%%%%%%%%%%%%%%%%%%
\section{Summary of the results\label{sec:results} }
%%%%%%%%%%%%%%%%%%%%%%%%%%%%%%%%%%%%%%%%%%%%%%%%%%%%%%

The main objective of this work is to amend the current formulation of high-energy factorization, which, in its present form, exhibits unstable evolution at next-to-leading order (NLO). The lack of convergence in the perturbative series has been extensively studied and is now understood to originate from large collinear logarithms, which stem from the absence of a proper time-ordering constraint. However, the commonly adopted resolution --- imposing a kinematic constraint to restore time ordering --- remains unsatisfactory from a formal standpoint, as it constitutes an \textit{ad hoc} fix that may compromise the control over subleading logarithms at any given order in perturbation theory.

A recent study \cite{Ducloue:2019ezk} has suggested that the problem arises from an inappropriate choice of rapidity variable, thereby casting doubt on the viability of the shockwave framework based on factorization in the $k^+$ variable. Nevertheless, a systematic framework for performing higher-order computations beyond the well-defined shockwave approach is still lacking.

In this work, we develop such a first-principles framework by:
\begin{enumerate}
\item retaining the shockwave formalism with $k^+$-ordering, along with its previously established results;
\item introducing a collinear factorization scale $\mu$, that allows to precisely define the rapidity phase space for high energy evolution;
\item exploiting the scheme dependence of rapidity factorization to reshuffle the large collinear logarithms (responsible for the instability of the rapidity evolution) from the evolution kernel to the initial condition and the impact factor.
\end{enumerate}

To next to leading order the factorized DIS cross-section, mediated by a longitudinally ($L$) or transversely ($T$) polarized virtual photon, reads schematically
\bea\label{eq:he-factorization-nlo}
& \sigma_{L/T}(x,Q^2) =  \, \left[H_{\rm LO}(Q^2) \otimes\, S_{\rm NLL}(\zeta)+\Delta H_{\rm NLO}(Q^2,\zeta/s) \otimes\, S_{\rm LL}(\zeta)\right]\,,
\eea
where $\zeta=2 \rho^+P^-$ denotes the rapidity factorization scale, $s=2 q^+P^-$. In addition, $S_{\rm LL}$ and $S_{\rm NLL}$ obey the LO (resumming leading logs (LL))  and NLO BK equations (resumming next-to-leading logs (NLL)), respectively. 
Explicit expressions $\Delta H_{\rm NLO}$ can be found in~\cite{Beuf:2017bpd,Hanninen:2017ddy} ; we provide in Appendix~\ref{app:HNLO} the formulas for the longitudinal DIS cross-section $\sigma_L(x,Q^2)$, which is slightly simpler. The LO term is standard textbook material and reads (see e.g.~\cite{Kovchegov:2012mbw})
\begin{align}
    H_{\rm LO}(Q^2)\otimes S_{\rm NLL}(\zeta)&=\sum_{i}\frac{4\alpha_{\rm em}e_i^2 N_cS_\perp}{\pi^2}\int\rmd^2\bx_{12}\int_0^1\rmd z_1 \ z_1(1-z_1)\bar Q^2K_0^2(\bar Qx_{12})(1-S_{12})\,,\label{eq:HLO-sigmaL}
\end{align}
with $\bar Q^2=z_1(1-z_1)Q^2$ and $S_\perp$ the transverse area of the target (we consider a homogeneous target in the transverse plane for simplicity). The LO contribution is expressed in terms of the dipole operator and defined as
\begin{align}
    S_{12}\equiv S(\bx_{12},\zeta)\equiv \frac{1}{N_c}\left\langle\textrm{Tr}\left[U_{\bx_1}U^\dagger_{\bx_2}\right]\right\rangle\,,
\end{align}
where $U_{\bx_1}$ and $U^\dagger_{\bx_2}$ denote infinite Wilson lines, which encode the color precession of the high-energy quark and antiquark forming the dipole as they traverse the target’s background field. Their explicit form is given in \eqn{eq:wilson-line}. Here, $\langle \cdots \rangle$ denotes the ensemble average over target configurations.

The $\zeta$-dependence arises from renormalization group evolution following the LL/NLL BK equation, as we shall see below. Throughout this paper, we use the shorthand notation $\bx_{ij}=\bx_i-\bx_j$ for the difference of two transverse coordinates.
The sum over $i$ runs over the $n_f$ light quark flavors with fractional electric charge $e_i$. 
 To leading logarithmic (LL) accuracy, the dipole operator $S(\bx_{12},\zeta) \equiv S_{12}$, obeys the BK equation

\beq \label{eq:lo-bk}
\frac{\del S_{12}}{\del \ln\zeta}  = \abar \int_3 K_{123} \left[S_{13}S_{32}-S_{12}\right]\,,
\eeq
where $\abar=\alpha_sN_c/\pi$ and the LO kernel reads 
\beq 
K_{123} =\frac{\bx_{12}^2}{\bx_{13}^2\bx_{32}^2}\,.
\eeq
We have used the shorthand notation for the transverse coordinate integrals 
\beq 
\int_3 \equiv \int \frac{\rmd^2\bx_3}{2\pi}\,.
\eeq
The NLL evolution equation was derived in \cite{Balitsky:2007feb} and takes the following form in the large $N_c$ limit
\begin{align}
\frac{\del S_{12}}{\del \ln\zeta} &=\abar K_{\rm NLO}\otimes  S_{12}\nn
&=  \abar \int_3 \, K_{123}^{\rm NLO}  \left[ S_{13} S_{32} - S_{12} \right]+ \frac{\abar^2}{2} \int_{3,4} K_{1234}  \left[ S_{13} S_{34} S_{42} - S_{13} S_{32} \right] + \frac{\abar^2 n_f }{2 N_c} \int_{3,4} K_f \,  S_{32} \left[ S_{14} - S_{13} \right]\,,\label{eq:NLO-BK}
\end{align}

where the various kernels have the following definitions:
\begin{align}
K_{123}^{\rm NLO} &= \frac{\bx_{12}^2}{\bx_{13}^2 \, \bx_{32}^2} \left[ 
1 + \frac{\alpha_s N_c}{4\pi} \left( 
\frac{\beta_0}{N_c} \ln (\bx_{12}^2 \mu_R^2) 
- \frac{\beta_0}{N_c} \frac{\bx_{13}^2 - \bx_{32}^2}{\bx_{12}^2} 
\ln \frac{\bx_{13}^2}{\bx_{32}^2} 
+ \frac{67}{9} - \frac{\pi^2}{3} - \frac{10}{9} \frac{n_f}{N_c} 
- 2\ln \frac{\bx_{13}^2}{\bx_{12}^2} \ln \frac{\bx_{32}^2}{\bx_{12}^2} 
\right) 
\right]\,,\label{eq:K123-NLO-def}\\
K_{1234}&= - \frac{2}{\bx_{34}^4} 
+ \left[ 
\frac{ \bx_{13}^2 \bx_{24}^2 + \bx_{14}^2 \bx_{32}^2 
- 4 \bx_{12}^2 \bx_{34}^2 }
{ \bx_{34}^4 (\bx_{13}^2 \bx_{24}^2 - \bx_{14}^2 \bx_{32}^2) } + \frac{ \bx_{12}^4 }{ \bx_{13}^2 \bx_{24}^2 (\bx_{13}^2 \bx_{24}^2 - \bx_{14}^2 \bx_{32}^2) } 
+ \frac{ \bx_{12}^2 }{ \bx_{13}^2 \bx_{24}^2 \bx_{34}^2 } 
\right]\ln \frac{ \bx_{13}^2 \bx_{24}^2 }{ \bx_{14}^2 \bx_{32}^2 }\,,\\
K_f &= \frac{2}{\bx_{34}^4} 
- \frac{ \bx_{14}^2 \bx_{32}^2 + \bx_{24}^2 \bx_{13}^2 - \bx_{12}^2 \bx_{34}^2 }
{ \bx_{34}^4 (\bx_{13}^2 \bx_{24}^2 - \bx_{14}^2 \bx_{32}^2) } 
\ln \frac{ \bx_{13}^2 \bx_{24}^2 }{ \bx_{14}^2 \bx_{32}^2 }\,.
\end{align}
In $K_{123}^{\rm NLO}$, $\beta_0=\frac{11}{3}N_c-\frac{2}{3}n_f$ and $\mu_R$ denotes the UV renormalization scale~\cite{Balitsky:2006wa,Kovchegov:2006vj}, not to be confused with the collinear scale $\mu$ discussed throughout this work. The double collinear logarithm -- the last term in $K_{123}$ given by \eqn{eq:K123-NLO-def},
\beq\label{eq:double-log} 
2\ln \frac{\bx_{13}^2}{\bx_{12}^2} \ln \frac{\bx_{32}^2}{\bx_{12}^2}\,,
\eeq
is known to generate numerical instabilities and therefore must be properly treated to ensure the convergence of the perturbative series and maintain control over theoretical uncertainties.

The transformation we propose, which leaves the cross section invariant, introduces an arbitrary collinear scale $\mu$, and results in the following modification of the dipole operator at next-to-leading order (NLO):
\beq \label{eq:composite-dipole}
S_{12} (\zeta)\quad \to \quad \bar S_{12}(\zeta,\mu^2) = S_{12}(\zeta) -  \abar \int_3 K_{123} \ln \br_{123}^2\mu^2 \left[ S_{13}S_{32}-S_{12}\right](\zeta)
\eeq
where $\br^2_{123}$ is a dynamical transverse scale that can be constructed from the the vectors $\bx_1$, $\bx_2$ and $\bx_3$ and whose precise form does not affect the leading collinear double logarithms.

We note that Eq.~\eqref{eq:composite-dipole} closely resembles the conformal dipole proposed in~\cite{Balitsky:2009xg}, with a key distinction: a factor of $1/2$, which plays a crucial role in fully canceling the double collinear logarithms. This subtle yet important difference, which directly links the coefficient of the logarithmic term $\ln \mu^2$ to the collinear divergences in the NLO BK equation~\eqn{eq:NLO-BK}, had not been identified previously and constitutes one of the central insights of our approach. In the context of BFKL, this is associated with the difference between the so-called DIS and $\gamma^\ast\gamma^\ast$ schemes. In the DIS scheme, the scale hierarchy $Q \gg Q_s$ breaks M\"obius invariance by spoiling the inversion symmetry $\bx \to \bx/\bx^2$. This reflects the fact that the dipole operator exhibits a different behavior at short and long distances, with saturation effects governing the latter regime. Drawing on the similarity with Ref.~\cite{Balitsky:2009xg}, we adopt their terminology and denote $\bar S$ as the  ``composite dipole".

Under the transformation \eqn{eq:composite-dipole},
the BK equation to NLO order takes the following form, 
\begin{align}
      \frac{\partial \bar S_{12}}{\partial \ln(\zeta)}
      &=\abar \tilde K_{\rm NLO}\otimes\bar S_{12}\label{eq:NLO-BK-comp-v2}+\abar^2\int_{3,4} K_{123}K_{134}\ln\left(\frac{\br_{423}^2\br_{134}^2}{\br_{123}^2\br_{124}^2}\right)(\bar S_{14}\bar S_{43}-\bar S_{13})\bar S_{32}\nonumber\\
    &-\abar^2\int_{3,4} K_{123}\left[K_{134}\ln\left(\frac{\br_{134}^2}{\br_{123}^2}\right)+ K_{324}\ln\left(\frac{\br_{124}^2}{\br_{143}^2}\right)+K_{124}\ln\left(\frac{\br_{123}^2}{\br_{124}^2}\right)\right](\bar S_{13}\bar S_{32}-\bar S_{12})\,.
\end{align}
We will show in detail, in particular in \secn{sec:mellin-space}, that this form of the NLO BK equation is free of large collinear logarithms. A convenient choice for the transverse vector $\br_{123}$ is 
\beq\label{eq:ms-scheme}
\br_{123}^2 = |\bx_{13}||\bx_{32}|\,,
\eeq
as it allows the cancelation of double logarithmic terms, individually, in both color structures $S^2 - S$ and $S^3 - S^2$. A formal justification for this  judicious choice will be presented in the next section. This scheme is, in fact, related to the kinematic constraint that ensures $k^- \le P^-$. The NLO BK equation can now be written explicitly as

\begin{empheq}[]{align}
   \frac{\partial \bar S_{12}}{\partial \ln(\zeta)}&=\abar\tilde K_{\rm NLO}\otimes\bar S_{12}\,+\frac{\abar^2}{2}\int_{3,4} K_{123}K_{324}\ln\left(\frac{\bx_{34}^2}{\bx_{14}^2}\right)(\bar S_{34}\bar S_{42}-\bar S_{32})\bar S_{13}\nonumber\\
    &+\frac{\abar^2}{2}\int_{3,4} K_{123}K_{134}\ln\left(\frac{\bx_{34}^2}{\bx_{24}^2}\right)(\bar S_{14}\bar S_{43}-\bar S_{13})\bar S_{32},\label{eq:composite-dipole-RG}
\end{empheq}
where $\tilde K_{\rm NLO}$ is the full NLO BK kernel \textit{without} the double collinear logarithm shown in the last term of Eq.~\eqref{eq:K123-NLO-def}.

Under the composite dipole transformation the NLO cross-section reads
\bea\label{eq:he-factorization-nlo-comp}
\sigma(x,Q^2) &= \bar H(Q^2,s/\zeta,\mu^2) \otimes {\bar S}(\zeta/\mu^2)\nn
& =  \, \bar H_{\rm LO}(Q^2) \,\otimes \bar S_{\rm NLL}(\zeta/\mu^2)+   \Delta \bar H_{\rm NLO}(Q^2,s/\zeta,\mu^2) \otimes \bar S_{\rm LL}(\zeta/\mu^2)+\cO(\abar^2)\,.
\eea
which differs from \eqn{eq:he-factorization-nlo-comp} by an additional dependence in the arbitrary collinear scale $\mu$. More explicitly, the composite dipole transformation amounts to adding the following term to Eq.~\eqref{eq:HNLO-sigmaL}:
\begin{align}\label{eq:barHNLO}
    \Delta\bar H_{\rm NLO}(Q^2,s/\zeta,\mu^2) \otimes \bar S_{\rm LL}(\zeta/\mu^2)&= \Delta H_{\rm NLO}(Q^2,\zeta/s)\otimes \bar S_{\rm LL}(\zeta)-\abar\sum_{i}\frac{4\alpha_{\rm em}e_i^2 N_cS_\perp}{\pi^2}\nonumber\\
    &\hspace{-1cm}\times\int\rmd^2\bx_{12}\int_0^1\rmd z_1 \ z_1(1-z_1)\bar Q^2K_0^2(\bar Qx_{12})\int\rmd^2\bx_3K_{123} \ln \br_{123}^2\mu^2 \left[ \bar S_{13}\bar S_{32}-\bar S_{12}\right](\zeta)\,.
\end{align}
It is interesting to note that such a counterterm in the NLO impact factor also appears when computing the DIS structure function with a rapidity regulator in the $k^-$ variable instead of the standard $k^+$ variable~\cite{Beuf:2025ggi}. Eqs.~\eqref{eq:composite-dipole}, \eqref{eq:composite-dipole-RG} and \eqref{eq:barHNLO} constitute the main results of this work.

Using the LO BK evolution equation \eqn{eq:lo-bk}, together with \eqn{eq:composite-dipole}, we can easily verify that 
\beq 
\frac{\del \bar S(\zeta,\mu^2)}{\del \ln \zeta} = -  \frac{\del\bar S(\zeta,\mu^2) }{\del \ln \mu^2} \,,\label{eq:pde-barS}
\eeq
which implies that $\bar S$ is function of $\zeta$ and $\mu$ through the scaling variable $\zeta/\mu^2$, that is, 
\beq\label{eq:S-scaling}
\bar S(\zeta,\mu^2) \equiv \bar S(\zeta/\mu^2)\,.
\eeq
This property, which \textit{a priori} holds at LO in $\alpha_s$, will be generalized to all orders in Section.~\ref{sec:comp-dip}. It entails three remarkable consequences:
\begin{enumerate}
    \item It allows one to choose the initial condition to be independent of the specific value of $\mu^2$, thus, 
\beq 
\bar S(Y_0=\ln(\zeta_0/\mu^2)) = \bar S(0) \,,
\eeq
for $\zeta_0=\mu^2$. In particular, it explains why one can choose $\mu$-independent initial condition, such as the McLerran-Venugopalan model~\cite{McLerran:1993ka,McLerran:1993ni} for the composite dipole.

    \item  It provides precise control over the definition of the rapidity variable entering the BK evolution. In particular, one can choose
\beq
\mu = Q, \quad \zeta = s 
\quad \Longrightarrow \quad 
Y = \ln\left(\frac{\zeta}{\mu^2}\right) = \ln\left(\frac{s}{Q^2}\right) = \ln\left(\frac{1}{x_{\rm Bj}}\right),
\eeq
which demonstrates that the change of the evolution rapidity variable can be systematically encoded within the shockwave formalism.

\item It offers guidance for building the composite dipole at NNLO, by enforcing the constraint Eq.~\eqref{eq:pde-barS} to be valid at NLO. Using this constraint together with Eq.~\eqref{eq:NLO-BK-comp-v2}, the composite dipole at $\mathcal{O}(\alpha_s^2)$ can be defined in terms of the original dipole as 
\begin{align}
    \bar S_{12}(\zeta,\mu^2)&= S_{12} -\abar K_{\rm NLO}\ln(\br_{1234}^2\mu^2)\otimes\bar S_{12}\nonumber\\
    &-\abar^2\int_{3,4} K_{123}K_{134}\left[\ln\left(\frac{\br_{134}^2}{\br_{123}^2}\right)\ln(\br_{1234}^2\mu^2)-\frac{1}{2}\ln^2(\br_{134}^2\mu^2)\right](\bar S_{14}\bar S_{43}-\bar S_{13})\bar S_{32}\nonumber\\
    &-\abar^2\int_{3,4} K_{123}K_{324}\left[\ln\left(\frac{\br_{324}^2}{\br_{123}^2}\right)\ln(\br_{1234}^2\mu^2)-\frac{1}{2}\ln^2(\br^2_{324}\mu^2)\right](\bar S_{34}\bar S_{42}-\bar S_{32})\bar S_{13}\nonumber\\
    &+\abar^2\int_{3,4} K_{123}K_{124}\left[\ln\left(\frac{\br_{124}^2}{\br_{123}^2}\right)\ln(\br_{1234}^2\mu^2)-\frac{1}{2}\ln^2(\br^2_{124}\mu^2)\right](\bar S_{14}\bar S_{42}-\bar S_{12})\,,\label{eq:composite-dipole-NNLO}
\end{align}

for some transverse coordinate $\br_{1234}$ defined in terms of $\bx_1,\bx_2,\bx_3$ and $\bx_4$. With this form of NNLO composite dipole, Eq.~\eqref{eq:composite-dipole-RG} also holds as an evolution equation in $Y$. Together with the recent calculation of the NNLO BK equation~\cite{brunello2025highenergyevolutionplanarqcd}, this result paves the way towards a consistence evaluation of the dipole operator at two-loops without collinear instability.

\end{enumerate}

%%%%%%%%%%%%%%%%%%%%%%%%%%%%%%%%%%%%%%%%%%%%%%%%%%%%%%
\section{The interplay between small-x and collinear logarithms  \label{sec:he-factorization} }
%%%%%%%%%%%%%%%%%%%%%%%%%%%%%%%%%%%%%%%%%%%%%%%%%%%%%%

High-energy factorization in deep inelastic scattering (DIS) is based on the separation between fast, $k^+>\rho^+$, long-lived quantum fluctuations, which are part of the photon impact factor $H$ and slow modes, $k^+<\rho^+$ effectively included in the multipole operators $S^{(m)}$, that encode the target properties:
\bea\label{eq:he-factorization}
& \sigma(x,Q^2) = \sum_{m=2}^\infty \,\prod_{i=2}^m\int_{\bx_i } H^{(m)}( Q^2,\{\bx_i\},\rho^+/q^+) \, S^{(m)}(\{\bx_i\},\rho^+P^-)\,.
\eea
(See Eqs.~\eqref{eq:HLO-sigmaL}-\eqref{eq:HNLO-sigmaL} for a concrete example of the structure given by Eq.~\eqref{eq:he-factorization} in the case of the longitudinal DIS structure function.)
Here $\rho^+$ is a rapidity factorization scale, in the frame where the photon is moving in the $+z$ direction with longitudinal momentum $q^+$ and the target is moving in the $-z$ direction with longitudinal momentum $P^-$. The sum runs over the number of high energy partons that cross the shockwave. 
At leading order the operator $S^{(2)}\equiv S$ is defined as a correlator of two infinite Wilson lines evaluated at the transverse positions of the quark and antiquark, $\bxi$ and $\bxii$:
\bea\label{eq:dipole}
& S(\bxi,\bxii) = \frac{1}{N_c}  \langle \rmTr \, U_{\bxi }U^\dag_{\bxii}\rangle\,.
\eea
with the infinite Wilson lines defined as
\beq\label{eq:wilson-line}
U_{\bx } \equiv \cP  \exp\left[ ig\int_{-\infty}^{+\infty} \rmd x^+ t\cdot A^-(x^+,\bx)\right]\,,
\eeq
with $t\cdot A^- \equiv t^a A^{-,a}$ where $t^a$ are the generator of $SU(N_c)$ in the fundamental representation.

Large rapidity logarithms are known to exponentiate  \cite{Caron-Huot:2015bja}
\beq\label{eq:evolution-op}
S^{(m)}(\{\bx_i\},\rho^+P^-)=  \exp\left[ \abar  K(\{\bx_i\},\abar) \ln \frac{\rho^+}{\rho^+_0}  \right]_{mn} S^{(n)}(\{\bx_i\},\rho_0^+P^-)\,,
\eeq
where the evolution operator $K$ -- in matrix form --  is a functional differential operator. We may equivalently write the differential form 
\beq
 \frac{\del S^{(m)}(\{\bx_i\},\zeta)}{\del \ln \zeta } =  \abar  K_{mn}(\{\bx_i\},\abar) S^{(n)}(\{\bx_i\},\zeta)\,.
\eeq 
which is essentially a formal rewriting of the Balitsky hierarchy as it results from the JIMWLK equation.
At leading order, only two components contribute to the evolution of the dipole operator $S^{(2)}$, namely, $K_{22}$ and $K_{23}$. The factorization scale $\zeta = 2\rho^+ P^-$ is bounded from above by the center-of-mass energy, $s = 2q^+ P^-$. However, there is an ambiguity on the lower bound, that may be $Q^2$, or some non-perturbative scale $\Lambda_{\rm QCD}^2$. This ambiguity is the origin of the collinear anomaly that results in the emergence of large collinear logs of these two scales order by order in perturbation theory.

The extension of the time integration domain of the Wilson line reflects the shockwave approximation, which decouples the $x^+$ integrals in the form factor from those in the target operator $S$. In momentum space, this corresponds to a separation in $k^-$, which follows from the ordering in $k^+$, modulo a transverse scale dependence. However, the condition $k^+ > \rho^+$ implies --- for real gluons, at least --- an upper bound on $k^-$ given by $k^- < k_\perp^2 / \rho^+$. This dynamical bound must remain below $P^-$, yet the factorization scheme does not account for this kinematic constraint, leading to collinear logarithms arising in the unphysical region $k_\perp^2 / \rho^+ > P^-$,
\beq
\ln \frac{k_\perp^2}{\rho^+ P^-},.
\eeq
This issue, which arises only in NLO and beyond, can be addressed in various ways. The most common remedy is to impose a kinematic constraint on the phase space of the quantum corrections in an {\it ad hoc} manner. In this paper, we develop a systematic method that exploits the freedom in the choice of subtraction scheme to address this issue order by order in perturbation theory within the shockwave approach.

As a first step, we revisit the one-loop computation. To do so, we take a step back and recognize that the problem described above originates from the extension of $x^+$ to infinity in the Wilson lines. To properly take the small-$x$ limit, we introduce an $x$-dependent dipole operator whose limiting behavior yields the dipole operator defined in Eq.~\eqref{eq:dipole}.

The one loop result for $S$ involving soft gluon emissions from the Wilson lines are both IR and UV divergent and are usually regulated with the help of a cutoff:
\beq
I = \int_0^{+\infty} \frac{\rmd k^+}{k^+} \to  \int_{\Lambda_0^+}^{\Lambda^+} \frac{\rmd k^+}{k^+}\,.
\eeq
However, while the upper limit indeed diverges for $S$ and requires a regulator, the lower bound should be regulated by the dynamical size of the shock wave. To see that, it is convenient to temporarily relax the small $x=0$ limit, to first carry out the integration over $k^+$, then take formally the limit $x \to 0$. The operator that enables this operation is defined as \cite{Boussarie:2021wkn,Boussarie:2020fpb}
\beq\label{eq:x-dipole}
 S(x,\bx_1,\bx_2)  &=& \frac{1}{N_c}\int_{-\infty}^{+\infty} \rmd x^+ \int_{-\infty}^{x^+}\rmd y^+  \, \rme^{i xP^- (x^+-y^+)}\frac{\del^2}{\del x^+\del y^+}  \rmtr \langle U_{\bx_1} (x^+,y^+) U_{\bx_2} ^\dag(x^+,y^+) \rangle \,,
\eeq
where we have introduced the finite Wilson line in the fundamental representation 
\beq\label{eq:finite-wilson-line}
U_{\bx } (x^+,y^+) \equiv [x^+,y^+]_{\bx}= \cP  \exp\left[ ig\int_{y^+}^{x^+} \rmd z^+ A^-(z^+,\bx)\right]\,.
\eeq
It can easily be checked that for $x=0$ the operator above reduces the dipole operator \eqn{eq:dipole}.

Using a rapidity regulator $\eta$ to regularize the $k^+\to +\infty$ divergence,  the one-loop correction to the above $x$-dependent dipole ($x$-DD) distribution reads (see Appendix~\ref{sec:evolutionequation} for a detailed derivation) 

\bea\label{eq:S2-PTE}
\Delta S_2(x,\bxi,\bxii) &= \, \abar (\rho^+)^\eta \int_{0}^{+\infty}  \rmd x' \, \int_{\bk,\bq,\bz}K(\bx_{13},\bx_{32},\bk,\bq)\,\nn & \times\int_0^{+\infty} \frac{\rmd k^+}{(k^+)^{1+\eta}} \delta\left(x'-x-\frac{(s\bk+(1-s)\bq)^2}{2P^-k^+}\right)\,  S_3 (x',\bxi,\bxii,\bxiii) \,.\nn
\eea
Here we have applied the partial twist expansion which is akin to a WKB approximation for the propagation of the radiated gluon inside the target \cite{Boussarie:2020fpb,Boussarie:2021wkn}.
The variable $s$ is arbitrary and will be chosen to be $s=1$. The dependence on this choice is expected to be subleading in power counting.

Here, we have used the contextual notation for the unintegrated LO BK kernel
\beq 
K(\bx_{13},\bx_{32},\bk,\bq) = (2\pi)^2   \frac{\bq\cdot \bk}{\bq^2 \bk^2 } \left(\rme^{-i\bk\cdot\bx_{13}}-\rme^{-i\bk\cdot\bx_{23}}\right) \left(\rme^{i\bq\cdot\bx_{13}}-\rme^{i\bq\cdot\bx_{23}}\right)\,,
\eeq
and 
\beq\label{eq:x-tripole}
S_3 (x',\bxi,\bxii,\bxiii)   &=& \frac{1}{N_c}\int_{-\infty}^{+\infty} \rmd x^+ \int_{-\infty}^{x^+}\rmd y^+  \, \rme^{i x'P^- (x^+-y^+)}\nn
 &\times& \frac{\del^2}{\del x^+\del y^+}  \rmtr \left[\langle t^ a U_{\bx_1} (x^+,y^+) t^b U_{\bx_2} ^\dag (x^+,y^+) \cU_{\bx_3}^{ab}(x^+,y^+)  \right]  \rangle \,.
\eeq
where 
\beq 
\cU_{\bx_3}^{ab}(x^+,y^+)  = \cP \exp\left[ ig \int_{y^+}^{x^+} \rmd z^+ T\cdot A^-(z^+,\bx_3)\right]^{ab}\,,
\eeq
is a Wilson-line in the adjoint representation.

For notational simplicity, we introduce the following shorthand 
\beq
 && S_2(x,\bxi,\bxii) \equiv  S_{12}(x)\,, \nn
 && S_3 (x',\bxi,\bxii,\bxiii)  \equiv S_{123}(x')\,, \nn
 && K(\bx_{13},\bx_{32},\bk,\bq)  \equiv K_{123}(\bk,\bq)\,.
\eeq
The LO BK kernel is recovered by integrating over the transverse momenta $\bq$ and $\bk$ as follows
\beq 
K_{123} = \int_{\bq,\bk} K_{123}(\bk,\bq) = \frac{\bx^2_{12}}{\bx^2_{13}\bx^2_{32}}\,.
\eeq
Again, similarly to the position space integrals the momentum space integral notation reads
\beq 
\int_{\bk}\equiv \int \frac{\rmd^2 \bk}{(2\pi)^2}\,.
\eeq
Hence, \eqn{eq:S2-PTE}, for $s=1$, becomes,
\bea\label{eq:S2-PTE-2}
\Delta S_{12}(x) &= \, \abar (\rho^+)^\eta \int_{0}^{+\infty}  \rmd x' \, \int_{\bk,\bq}\int_3K_{123}(\bk,\bq)\int_0^{+\infty} \frac{\rmd k^+}{(k^+)^{1+\eta}} \delta\left(x'-x-\frac{\bk^2}{2P^-k^+}\right)\,  S_{123} (x') \,.
\eea

Note that neglecting the $x$ and $\bk^2/2k^+$ terms in the Dirac delta function, we can readily integrate over $\bk$ and $\bq$ and as a result we recover the BK kernel: 
\beq\label{eq:naive-smallx}
\Delta S_{12}(0)&=& \, \abar (\rho^+)^\eta \, \int_0^{+\infty} \frac{\rmd k^+}{(k^+)^{1+\eta}}\int_{3}\frac{\bx_{12}^2}{\bx_{13}^2\bx_{32}^2} \,  S_{123} (0)\,.
\eeq
This result is the standard shockwave approximation, however, it does not fully reflect the proper small $x$ limit which should include an additional subleading term which emerges when keeping $\bk^2/2k^+$ as we shall shortly see.  

Taking the $x\to 0$ limit of \eqn{eq:S2-PTE}, we have
\beq
\Delta S(x=0) =\lim_{x\to 0} \Delta S(x) &=& \lim_{x\to 0} \, (\rho^+)^\eta \int_{0}^{+\infty}  \rmd x' \,  \int_{\bk,\bq}\int_3K_{123}(\bk,\bq)\int_0^{+\infty} \frac{\rmd k^+}{(k^+)^{1+\eta}} \delta\left(x'-x-\frac{\bk^2}{2P^-k^+}\right)\,  S^{(3)} (x') \,,\nn
&=&  (\rho^+)^\eta\int_{0}^{+\infty}  \rmd x'  \int_{\bk,\bq}\int_3K_{123}(\bk,\bq)\int_0^{+\infty} \frac{\rmd k^+}{(k^+)^{1+\eta}} \delta\left(x'-\frac{\bk^2}{2k^+ P^-}\right)\,  S^{(3)}(x')\,,\\
&=&  \int_{\bk,\bq}\int_3K_{123}(\bk,\bq) \left(\frac{2\rho^+P^-}{\bk^2}\right)^\eta\int_{0}^{+\infty}  \frac{ \rmd x' }{(x')^{1-\eta}} \,  S^{(3)}(x')\,.
\eeq
The delta function in $x$ can be understood as encoding the kinematic constraint $xk^- < k^- < P^-$ which would be imposed by hand in other \textit{ad hoc} schemes.

Since we are interested in the limit $\eta \to 0$, it is convenient to expand the integrand as  
\beq
  \frac{1}{(x')^{\,1-\eta}}
  = \frac{1}{\eta}\,\delta(x') 
  + \left[\frac{1}{x'}\right]_+ 
  + \eta \left[\frac{\ln x'}{x'}\right]_+ 
  + \mathcal{O}(\eta^2)\,,
\eeq
where the plus distributions are defined through their action on a smooth test function $f(x')$.  
In particular,
\beq
  \int_0^1 \! dx'\, \left[\frac{1}{x'}\right]_+ f(x')
  = \int_0^1 \! dx'\, \frac{f(x') - f(0)}{x'}\,.
\eeq
Thus, we  find
\beq
\Delta S^{(1)}(x=0) &\simeq &   \left(\frac{2\rho^+P^-}{\bk^2}\right)^\eta\int_{0}^{+\infty}   \rmd x'  \left\{\frac{1}{\eta}\delta(x') + \left[\frac{1}{x'}\right]_+\right\}  \,  S^{(3)}(x')\,\nn
&\simeq&   \left(1+ \eta \ln \frac{2\rho^+P^-}{\bk^2}\right) \left\{\frac{1}{\eta}S^{(3)}(0)  +  \int_{-\infty}^{+\infty}    \frac{\rmd x'}{x'}(S(x')-S(0)) \right\}  \,  \,\nn
&\simeq& \left(\frac{1}{\eta}+\ln \frac{2\rho^+P^-}{\bk^2} \right)S^{(3)}(0)  + \int_{0}^{+\infty}    \frac{\rmd x'}{x'}(S^{(3)}(x')-S^{(3)}(0))   \,  \,\nn
&\simeq& \left( \frac{1}{\eta}+\ln \frac{2\rho^+P^-}{\bk^2} \right)\, S^{(3)}(x=0)  +... \,  \label{eq:S1loop-ms}
\eeq
The ellipsis denotes terms that are regular as $x\to 0$. 

Interestingly, we obtain a logarithm of the ratio of the  factorization scale and the dynamical scale $k^+=\bk^2/2P^-$. This collinear log was missed from the standard shockwave approach and can be shown to be related to the so-called kinematic constraint. Although this may be expected when $k^-$ ordering is enforced, what is less trivial is that the transverse logarithm multiplies the dipole operator evaluated at precisely $x=0$. In addition, we obtain additional non-dipole operators that encode physics at finite $x$ which we shall discard. This demonstrates that the kinematic constraint does not spoil the high energy operator product expansion (OPE) formulated in terms of operators evaluated at $x=0$. As such, there should be a way to formulate this OPE without formally breaking it by an ad-hoc modification of the kernel for the high-energy evolution, as currently done in the literature.

To identify the proper OPE formulation of the kinematic constraint,  we first relate the result Eq.~\eqref{eq:S1loop-ms} to the BK equation written in coordinate space. We first
introduce a separation scale $\rho_0^+\ll \rho^+$ and write 
 \beq
\Delta S(x=0) 
&\simeq& \left( \frac{1}{\eta}+\ln \frac{\rho^+}{\rho_0^+} - \ln\frac {\bk^2}{2\rho_0^+P^-} \right)\, S^{(3)}(x=0)  +{\rm const.} \,,\\
&\simeq& \Delta S_{\rm BK}(x=0) - \ln \frac{\bk^2}{2\rho_0^+P^-} \, S^{(3)}(x=0)  +{\rm const.} \,.
\eeq
More precisely, setting 
\beq
\fscal^2 = 2\rho^+_0P^- \,,
\eeq
we have

  \beq
\Delta S_{12}(x=0) 
&\simeq& \Delta S^{\rm BK}_{12}(x=0) - \abar\int_{3} L_{123}(\fscal^2) \, S_{123}^{(3)}(x=0)   +{\rm const.} \, ,
\eeq
where
\beq\label{eq:BK-remainder}
L_{123}(\fscal^2) =\frac{(2\pi)^2}{2} \int_{\bq,\bk}  \frac{\bq\cdot \bk}{\bq^2 \bk^2 } \left(\rme^{-i\bk\cdot\bx_{13}}-\rme^{-i\bk\cdot\bx_{23}}\right) \left(\rme^{i\bq\cdot\bx_{13}}-\rme^{i\bq\cdot\bx_{23}}\right) \left( \ln\frac {\bk^2}{\fscal^2}  +\ln\frac {\bq^2}{\fscal^2}   \right)\,,
\eeq
which can be integrated using  (cf. Appendix.~\ref{app:integrals})
\beq
\int_{\bq}  \frac{\bq^i }{\bq^2 }  \left(\rme^{i\bq\cdot\bx_{13}}-\rme^{i\bq\cdot\bx_{23}}\right)
&=& \frac{i}{2\pi}  \left( \frac{\bx_{13}^i}{\bx_{13}^2}-\frac{\bx_{23}^i}{\bx_{23}^2}\right)\,,
\eeq
and 
\beq
 \int_{\bk}  \frac{\bk^i }{\bk^2 }  \left(\rme^{-i\bk\cdot\bx_{13}}-\rme^{-i\bk\cdot\bx_{23}}\right)\ln\frac {\bk^2}{\fscal^2}
&=&  -\frac{i}{(2\pi)}  \left[ \left( \ln \frac { |\bx_{13}||\bx_{23}|}{\fscal^2}  +2 (\gamma_{E}-\ln(2)) \right) \left( \frac{\bx_{13}^i}{\bx_{13}^2}-\frac{\bx_{23}^i}{\bx_{23}^2}\right)+\left( \frac{\bx_{13}^i}{\bx_{13}^2}+\frac{\bx_{23}^i}{\bx_{23}^2}\right)  \ln\frac{|\bx_{13}|}{|\bx_{23}|}  \right] \,.\nn
\eeq
We thus obtain 
\beq
L_{123}(\fscal^2)& =&  \frac{\bx_{12}^2}{\bx_{13}^2\bx_{23}^2} \ln ( |\bx_{13}||\bx_{23}| \fscal^2)  \,  +\left( \frac{1}{\bx_{13}^2}-\frac{1}{\bx_{23}^2}\right)  \ln\frac{|\bx_{13}|}{|\bx_{23}|}  \,\nn
 & =& \frac{\bx_{12}^2}{\bx_{13}^2\bx_{23}^2} \left[\ln \left(\frac{ |\bx_{13}||\bx_{23}|}{|\bx_{12}|\mu^{-1}}\right) + \ln   |\bx_{12}|\fscal \right] \,  +\left( \frac{1}{\bx_{13}^2}-\frac{1}{\bx_{23}^2}\right)  \ln\frac{|\bx_{13}|}{|\bx_{23}|}   \,.\nn\label{eq:S12-oneloop}
\eeq
where we have absorbed the additive constant in a redefinition of the factorization scale 
\beq 
\fscal^2 \to \fscal^2 \rme^{-2 (\gamma_{E}-\ln(2)) }\,.
\eeq

To sum up, the one loop formula in coordinate space for the dipole operator reads
\beq\label{eq:NLO-generalized dipole}
\Delta S_{12} &= & \abar \ln \frac{\rho^+P^-}{\fscal^2} \int_3\frac{\bx^2_{12}}{\bx^2_{13} \bx^2_{32}} \left(S_{13}S_{32}-S_{12}\right) \nn
&- &  \abar  \int \frac{\bx^2_{12}}{\bx^2_{13}\bx^2_{32}} \left( \ln \frac{|\bx_{13}||\bx_{32}|\fscal}{|\bx_{12}|}+ \ln |\bx_{12}|\fscal+\frac{\bx^2_{23}-\bx^2_{13}}{\bx^2_{12}} \ln \frac{|\bx_{13}|}{|\bx_{32}|}\right)\left(S_{13}S_{32}-S_{12}\right) \,.
\eeq
 We shall refer to the second term as the collinearly anomalous term. In this term, we have split the logarithm into two terms to highlight the connection with the ``conformal dipole" formulation \cite{Balitsky:2009xg,Balitsky:2010ze}. The first logarithm
\beq
  \ln \frac{|\bx_{13}||\bx_{32}|\fscal}{|\bx_{12}|} \,,
\eeq
corresponds to the  ``conformal dipole" term. The second one, 
 \beq
\ln |\bx_{12}|\fscal \,,
\eeq
breaks conformal invariance. It is known that the conformal dipole prescription cancels the explicit double logarithms in the $K_{123}^{\rm NLO}$ kernel (cf Eq.~\eqref{eq:K123-NLO-def}) responsible for the instabilities in the rapidity evolution of the NLO BK equation. However, it introduces a $\ln |\bx_{12}|$ term that generates instabilities of its own~\cite{Lappi:2015fma}, because it generates a double collinear logarithm ``hidden" in the $K_{1234}$ kernel. We anticipate that our additional term will resolve this issue.

%%%%%%%%%%%%%%%%%%%%%%%%%%%%%%%%%%%%%%%%%%%%%%%%%%%%%%
\section{Rapidity factorization scheme transformation\label{sec:comp-dip}}
%%%%%%%%%%%%%%%%%%%%%%%%%%%%%%%%%%%%%%%%%%%%%%%%%%%%%%

We turn now to a more formal discussion of the high energy OPE in terms of a our composite dipole, in close analogy with that of the conformal dipole introduced by Balitksy and Chirilli.
Because the subtraction of the rapidity singularity is scheme-dependent --- up to terms that may depend on the transverse coordinates --- the second term in \eqref{eq:S12-oneloop} can be viewed as a different choice of subtraction scheme~\cite{Caron-Huot:2015bja}. At one-loop order, this new subtraction scheme amounts to a redefinition of the dipole operator into a composite operator:
\beq\label{eq:NLO-composite-dipole}
&&\bar S_{12}\equiv S_{12}+ \abar  \int_3\frac{\bx^2_{12}}{\bx^2_{13}\bx^2_{32}}  \left( \ln \frac{1}{|\bx_{13}||\bx_{32}| \fscal^2}+\frac{\bx^2_{13}-\bx^2_{23}}{\bx^2_{12}} \ln \frac{|\bx_{13}|}{|\bx_{32}|}\right)\left(S_{13}S_{32}-S_{12}\right) +\mathcal{O}(\alpha_s^2)
\eeq
Interestingly, the second term coincides with the result recently derived in the $k^-$ scheme \cite{Beuf:2025ggi}, pointing to an equivalence between our rotated $k^+$ scheme and the direct choice of $k^-$ as the ordering variable. 

This non-linear transformation may seem obscure at first glance, but it can be naturally understood as a change of vector basis. Formally, it can be written to all orders as:
\beq
& & \bar S^{(m)}(\{ \bx_i \}, \zeta,\mu^2)=  \exp\left[ - \abar L (\{\bx_i\},\abar,\mu^2) \right]_{mn} S^{(n)}(\{ \bx_i\}, \zeta)\,.\label{eq:allorder-composite-dipole}
\eeq
As a particular case, we have for the dipole operator at NLO (recall our shorthand notation, $S_{12}=S^{(2)}(\{\bx_1,\bx_2\},\zeta)$)
\beq\label{eq:compositedipole-general}
& & \bar S_{12}(\zeta,\mu^2)= S_{12} (\zeta) -\abar  \int_3 L_{123}(\mu^2)  \left(S_{13}S_{32}-S_{12}\right)\,,\quad L_{123}\equiv K_{123}  \left( \ln|\bx_{13}||\bx_{32}| \fscal^2+\frac{\bx^2_{32}-\bx^2_{13}}{\bx^2_{12}} \ln \frac{|\bx_{13}|}{|\bx_{32}|}\right)\,,
\eeq
after expanding to first order in $\alpha_s$ the exponential in Eq.~\eqref{eq:allorder-composite-dipole} and identifying the result to Eq.~\eqref{eq:NLO-composite-dipole}.
As a result, both the evolution operator and the impact factor transform under this change of basis. Formally, the composite operators evolve according to (compare to Eq.~\eqref{eq:evolution-op})
\beq
 \bar S^{(m)}(\{\bx_i\},\zeta,\mu^2)=  \exp\left[\abar  \bar K(\{\bx_i\},\abar,\mu^2) \ln \frac{\zeta}{\zeta_0}  \right]_{mn} \bar S^{(n)}(\{\bx_i\},\zeta_0,\mu^2)
\eeq
where 
\beq\label{eq:kernel-transformation}
  \bar K  = \rme^{ - \abar L }  K\,  \rme^{  + \abar L}  =  K  + \abar [ K,L]+\frac{\abar^2}{2!}[[K,L],L]+\mathcal{O}(\abar^3)
\eeq
This result shows that the composite dipole evolves at NLO according to the NLO BK equation amended by an extra term given by the commutator between the kernels $K$ and $L$. The explicit computation of this commutator in coordinate space will be performed in Sec.~\ref{sec:sudakov}~\footnote{We note that a similar strategy was used in the context of the BFKL equation in position space\cite{Fadin:2006ha,Fadin:2007de}}.

\textit{A priori}, the composite operators depend on the factorization scale $\zeta=\rho^+P^-$ and the collinear scale $\mu^2$. A remarkable feature of the NLO composite dipole is the following relation:
\begin{align}
\frac{\partial \bar S_{12}}{\partial \ln\mu^2}=-\frac{\partial \bar S_{12}}{\partial \ln\zeta}+\mathcal{O}(\abar^2)\label{eq:lo-derivative-relation}
\end{align}
This becomes evident by taking the $\ln\mu^2$ derivative of Eq.~\eqref{eq:NLO-composite-dipole}, which exactly reproduces the LO BK equation given in Eq.~\eqref{eq:lo-bk}. The identity in Eq.~\eqref{eq:lo-derivative-relation} implies that, at least to LO in $\alpha_s$, $\bar S_{12}$ depends only on the ratio $\zeta/\mu^2$. This property allows us to relate our composite dipole framework to the choice of evolution variable in high-energy evolution. For $\mu^2\sim Q^2$ and $\rho^+\sim q^+$, we find $\ln(\zeta/\mu^2)\sim \ln(1/x_{\rm Bj})$, and therefore the composite dipole obeys the same evolution equation whether expressed as a function of $\ln(1/x_{\rm Bj})$ (target rapidity ordering) or $\ln(1/\zeta)$ (projectile rapidity ordering).

We now aim to impose this property to all orders in $\alpha_s$. Specifically, we require the composite operators defined by the general transformation Eq.,\eqref{eq:allorder-composite-dipole} to satisfy the constraint
\begin{align}
\frac{\partial \bar S^{(m)}}{\partial \ln\mu^2}+\frac{\partial \bar S^{(m)}}{\partial \ln\zeta}&=0,,\label{eq:derivative-relation}
\end{align}
or equivalently, $\bar S^{(m)}=\bar S^{(m)}({\bx_i},\zeta/\mu^2)$. This constraint serves as our guiding principle for constructing the composite dipole order by order in perturbation theory, given the perturbative expansion of the kernel $K$. We note, however, that one could instead impose a different constraint than Eq.~\eqref{eq:derivative-relation}, for example $\frac{1}{2}\frac{\partial \bar S^{(m)}}{\partial \ln\mu^2}+\frac{\partial \bar S^{(m)}}{\partial \ln\zeta}=0$, which corresponds to a symmetric ordering $\ln(\rho^+/\mu)$. This choice is related to the conformal dipole introduced in~\cite{Balitsky:2009xg}. The derivative of $\bar S^{(m)}$ with respect to $\ln\zeta$ follows from the definition of the kernel $\bar K$,
\begin{align}
    \frac{\partial \bar S^{(m)}}{\partial \ln\zeta}&=\abar  \bar K_{mn}(\{\bx_i\},\abar,\mu^2) \bar S^{(n)}\,,
\end{align}
and the derivative with respect to $\ln\mu^2$ is obtained from the definition of the composite operators,
\begin{align}
\frac{\partial\bar{S}^{(m)}}{\partial\ln\mu^{2}} & =-\abar\int_{0}^{1}\!{\rm d}u\,{\rm e}^{-u\abar L}\frac{\partial L}{\partial\ln\mu^{2}}{\rm e}^{u\abar L}\bar{S}^{(m)}\,,
\end{align}
where we have used \eqn{eq:allorder-composite-dipole} together with the identity, 
\beq \label{eq:dexpL}
 \frac{\rmd }{\rmd \ln \mu^2} \rme^{-\abar L(\mu)} = -\abar  \int_0^1 \rmd u \rme^{-(1-u) \abar L(\mu)}\frac{\rmd L(\mu)}{\rmd \ln \mu^2}   \rme^{-u \abar L(\mu)}\,,
\eeq
whose derivation is presented in Appendix~\ref{app:Kbar-derivative}. 

Imposing \eqn{eq:derivative-relation} for all $\bar S^{(m)}$, we find the identity 
\begin{align}
   \bar{K}= \int_{0}^{1}\!{\rm d}u\,{\rm e}^{-u\abar L}\frac{\partial L}{\partial\ln\mu^{2}}{\rm e}^{u\abar L}.\label{eq:cond}
\end{align}
As a consequence of this identity, the kernel $\bar K$ for the high-energy evolution of the composite dipole is independent of $\mu^2$, to all orders in $\alpha_s$. Indeed,
\begin{align}
    \frac{\partial\bar{K}}{\partial\ln\mu^{2}} & =\abar\int_{0}^{1}\!{\rm d}u\left[\bar{K},{\rm e}^{-u\abar L}\frac{\partial L}{\partial\ln\mu^{2}}{\rm e}^{u\abar L}\right]=0\label{eq:Kbar-derivative}
\end{align}
This property is a consequence of the linear relation Eq.\,\eqref{eq:derivative-relation} between the $\ln\zeta$ and $\ln\mu^2$ derivatives imposed on the composite dipole. It means that to all orders in $\alpha_s$, the kernel of the Balitsky hierarchy for the composite operators has no explicit $\mu^2$ dependence, a feature that we shall explicitly observe at NLO for the high-energy evolution of the composite dipole in Sec.~\ref{sec:sudakov}.

The condition in Eq.~\eqref{eq:cond} also imposes strong constraints on the composite dipole, allowing us to determine the form of the transformation given by Eq.~\eqref{eq:allorder-composite-dipole} order by order in perturbation theory, provided the BK kernel for the ``bare" dipole is known at the corresponding order. Let us illustrate this by explicit formal manipulations in order to obtain the $\mathcal{O}(\alpha_s^2)$ in Eq.~\eqref{eq:NLO-composite-dipole}. Up to $\mathcal{O}(\alpha_s^2)$ corrections, if we write
\begin{equation}
K=K_{1}+\bar\alpha_{s}K_{2}\,,\quad L= L_1+\bar\alpha_sL_2=L_{10}+L_{11}\ln\mu^2+\bar\alpha_s(L_{20}+L_{21}\ln\mu^2+L_{22}\ln^2\mu^2)\,,
\end{equation}
where $K_1$ and $K_2$ are the $\mathcal{O}(\alpha_s)$ and $\mathcal{O}(\alpha_s^2)$ BK kernel respectively, then the condition~Eq.~\eqref{eq:cond} reads
\begin{align} \label{eq:KNLO-formal}
 \bar K_{\rm NLO}= \frac{\partial L_{1}}{\partial\ln\mu^{2}}+\bar\alpha_s\frac{\partial L_{2}}{\partial\ln\mu^{2}}+\frac{1}{2}\bar\alpha_s\left[\frac{\partial L_{1}}{\partial\ln\mu^{2}},L_{1}\right] =K_{1}+\bar\alpha_sK_{2}+\bar{\alpha}_{s}[K_{1},L_{1}]\,.
\end{align}

Identifying term by term and solving order by order the corresponding elementary differential equation, we get 
\begin{align}
    L_1&= L_{10}+K_1\ln\mu^2\,,\label{eq:L1-formal}\\
    L_2&= L_{20}+\left\{K_2+\frac{1}{2}[K_1,L_{10}]\right\}\ln\mu^2\,.\label{eq:L2-formal}
\end{align}
Given the $\mathcal{O}(\alpha_s^3)$ BK kernel $K_3$, recently computed in \cite{brunello2025highenergyevolutionplanarqcd}, it is straightforward to obtain the formal expression for $L_3$, the $\mathcal{O}(\alpha_s^3)$ term in the expansion of $L$,
\begin{equation}
L_{3}=L_{30}+\left\{K_{3}+\frac{1}{2}[K_{1},L_{20}]+\frac{1}{2}[K_{2},L_{10}]+\frac{1}{12}[[K_{1},L_{10}],L_{10}]\right\}\ln\mu^{2}+\frac{1}{12}[K_{1},[K_{1},L_{10}]]\ln^{2}\mu^{2},
\end{equation}
and so forth for the higher order terms. In particular, using \eqn{eq:kernel-transformation} and the above relations we can provide the modification of the BK kernel at NNLO, i.e., $K_{\rm NNLO}=K_1 +\abar K_2+\abar^3 K_3$, 
\beq\label{eq:KNNLO-formal}
 \bar K_{\rm NNLO} =K_{\rm NNLO}+ \bar{\alpha}_{s}[K_{1},L_{10}]+\abar^2 [K_2,L_{10}]+\abar^2 [K_1,L_{20}]+\frac{\abar^2}{2}[[K_1,L_{10}],L_{10}]\,.
\eeq
The two relations in \eqn{eq:L1-formal} depend on arbitrary “constants” $L_{10}$ and $L_{20}$, which reflect the freedom to choose the transverse coordinate scale that renders the argument of $\ln\mu^2$ dimensionless. In Sec.~\ref{sec:mellin-space}, we will provide another argument based on the Mellin space representation of the linearized NLO BK equation for the composite dipole demonstrating that indeed, what truly matters for canceling the collinear logarithms in the NLO equation is the coefficient in front of the $\ln\mu^2$ dependence and not the precise form of the argument of the logarithm in the constant $L_{10}$. To connect with the composite dipole as defined in Eq.~\eqref{eq:composite-dipole} or Eq.~\eqref{eq:NLO-composite-dipole}, we have $L_{10}S_{12}=\int_3 K_{123}\ln(\br_{123}^2)(S_{13}S_{32}-S_{12})$ --- the rule for applying $L_{10}$ on more complicated operators involving more than two Wilson lines follows, in the large $N_c$ limit, from a ``Leibniz rule", i.e.~$L_{10}[S_{ab}S_{cd}]=(L_{10}S_{ab})S_{cd}+S_{ab}(L_{10}S_{cd})$. We thus find that imposing the constraint Eq.~\eqref{eq:cond} naturally leads to the form of the composite dipole as it emerges from the NLO calculation of the generalized dipole performed in Sec.~\ref{sec:he-factorization}. Since we have some freedom in choosing $L_{10}$, the simplest choice for $L_{10}$ could be something like
\beq
 L_{10} = \int_3 K_{123} \ln \bx^2_{12}\,.\label{eq:simple-composite-dipole}
\eeq
Yet, as we shall see in Sec.~\ref{sec:sudakov}, this choice does not lead to a simple form for the NLL BK equation satisfied by the composite dipole. We will explore other possibilities for the transverse coordinate scale accompanying $\mu$ in this section.

Thus, even without performing the two-loop calculation of the generalized dipole discussed in Sec.~\ref{sec:he-factorization}, the general structure of the change of basis, together with the condition we impose to extend the simple scaling property of the composite dipole $\bar S_{12}(\zeta/\mu^2)$ to all orders, determines the form of this composite dipole such that the resulting high energy evolution equation is collinear safe. To get  the order $\mathcal{O}(\alpha_s^2)$ composite dipole, one simply needs to take Eq.~\eqref{eq:L2-formal} with some constant operator $L_{20}$ and expand the exponential in Eq.~\eqref{eq:allorder-composite-dipole} up to order $\alpha_s^2$. The result of this manipulation is Eq.~\eqref{eq:composite-dipole-NNLO} displayed in the summary section.

Let us finally conclude this section by a brief discussion of the impact factor which is also affected by the change of basis corresponding to the transformation Eq.~\eqref{eq:allorder-composite-dipole}. Since the physical cross section $\sigma(x,Q^2)$ must be independent of the chosen basis, the impact factor must transform accordingly:
\beq
\bar H^{(n)}( Q^2,\{\bx_i\},\rho^+/q^+,\mu^2)  =    H^{(m)}( Q^2,\{\bx_i\},\rho^+/q^+) \exp\left[\abar L (\{\bx_{i}\},\abar,\mu^2) \right]_{mn}
\eeq
Hence, at NLO the cross-section takes the form 
\bea\label{eq:he-factorization-NLO}
 \sigma_{\rm NLO}(x,Q^2) &= H_{\rm NLO}(\rho^+/q^+) \, (1+ \abar L_1)\, \exp\left[\abar(K_1 + \abar [K_1,L_1] )\ln \frac{\rho^+}{\rho^+_0}  \right]\bar S (\rho_0^+P^-)+\cO(\abar^2)\,,\\
&=\bar H_{\rm NLO}(\rho^+/q^+,\mu^2)\exp\left[ \abar(K_1+ \abar [K_1,L_1] )\ln \frac{\rho^+}{\rho^+_0}  \right]\bar S (\rho_0^+P^-)+\cO(\abar^2)
\eea
where $\bar H_{\rm NLO}(\rho^+/q^+,\mu^2)= H_{\rm NLO}(\rho^+/q^+) \, (1+ \abar L_1(\mu^2) )=H_{\rm NLO}(\rho^+/q^+)+\bar\alpha_s H_{\rm LO}L_1(\mu^2)$. The explicit form for $\bar H_{\rm NLO}$ (convoluted with the evolved composite dipole) in the case of the longitudinal structure function is shown in Eq.~\eqref{eq:barHNLO} of the summary section. Hence, although the $\mu$-dependent terms in the operator $L$ cancel out in the evolution, its exact form remains essential for determining the impact factor within this scheme. In the next sections, we will examine the effect of this transformation on the collinear double logarithms that arise at NLO.

%%%%%%%%%%%%%%%%%%%%%%%%%%%%%%%%%%%%%%%%%%%%%%%%%%%%%%
\section{Heuristic Discussion of the Anomalous Collinear Logarithms \label{sec:HNLO}}
%%%%%%%%%%%%%%%%%%%%%%%%%%%%%%%%%%%%%%%%%%%%%%%%%%%%%%
It is not obvious from the structure of Eq.\eqref{eq:NLO-BK} that the pathological terms appearing in high energy factorization beyond leading order will indeed cancel out. As briefly summarized in the introduction, the pathology of naive high energy factorization manifests itself in two places: \begin{itemize}
    \item[(i)] at the level of the impact factors computed order by order in perturbation theory, which become artificially large (such that the cross-section can even become negative) if the factorization scheme ``oversubtracts" the high energy logarithms in the fixed order result,
    \item[(ii)] in the renormalization group equation resumming these high energy logarithms beyond leading order which displays large collinear logarithms in such naive factorization schemes.
\end{itemize}
How a composite dipole formulation based on Eq.~\eqref{eq:NLO-BK} solves the issue (ii) will be the subject of the following sections. Here, we would like to address how it also generically addresses the issue (i). Instead of discussing the exact NLO impact factor for the DIS structure functions (given in Sec.~\ref{sec:results} for longitudinally polarized virtual photons), we shall proceed with a heuristic argument based on a double logarithmic approximation of both the hard coefficient function and the dipole operator.

As first realized in~\cite{Beuf:2014uia} in the context of the NLO impact factors for the DIS structure functions at small $x$, the negativity problem of NLO cross-sections in high energy factorization comes from the fact that the fixed order NLO results knows about life-time ordering while the LO BK equation formulated with the evolution variable of the projectile does not. In DIS, the lifetime of the radiated gluon in the color dipole picture must be smaller than that of the virtual photon creating the $q\bar q$ pair. It is then not obvious at all to see how formula Eq.~\eqref{eq:NLO-BK} knows about lifetime ordering. To convince ourselves that this is indeed the case, let us analyze the double logarithmic limit of gluon radiation at the level of the cross section \eqn{eq:he-factorization}.
At NLO the impact factor knows about one side of the ordered sequence; the gluon formation time is shorter than the photon decay time:
\beq
\frac{k^+}{\bk^2 } \, \ll \,  \frac{q^+}{Q^2 }    \quad  \implies \quad k^+ \ll  \left(\frac{\bk^2}{Q^2 }\right) q^+
\eeq
This upper bound on $k^+$ is effectively encoded inside the NLO light-cone wave function: mathematically, the Bessel function $K_0(QX_R)$ in Eq.~\eqref{eq:HNLO-sigmaL} suppress the phase-space for high-energy evolution when $k^+\gg q^+/(\bx_{13}^2Q^2)\sim (\bk^2/Q^2)q^+$.
Applying the above constraint to the hard coefficient function at NLO should generate the following term 
\beq
 H_{\rm NLO}(Q^ 2,\rho^+/q^+) &\approx &\bar\alpha_s \int^{Q^2}_{Q_0^2} \frac{\rmd \bk^2}{\bk^2} \int^{\left(\frac{\bk^2}{Q^2 }\right) q^+}_{\rho^+} \frac{\rmd k^+}{k^+} \,\\
 &=& \bar\alpha_s\left[ \ln\frac{Q^2}{Q_0^2}\ln\frac{q^+}{\rho^+} - \frac{1}{2} \ln^2\frac{Q^2}{Q_0^2}\right]\,,\label{eq:C1-DL}
\eeq
where $Q_0\ll Q^2$ is the characteristic (non-perturbative) scale of the target. The first term is \textit{a priori} a small correction since we typically have $\rho^+\sim q^+$: it features the double logarithm present in a single step of a  \textit{projectile} ($k^+$) ordered high-energy evolution between $\rho^+$ and $q^+$. On the other hand, the second term in this expression is a pure NLO correction from the point of view of the resummation of high-energy logarithms ; yet it can be large when $Q^2\gg Q_0^2$ such that $H_{\rm NLO}$ becomes negative. The dipole operator on the other hand yields
\beq
 S(|\br|\sim Q^{-1},\rho^+ )&\approx& \bar\alpha_s \int^{Q^2}_{Q_0^2} \frac{\rmd \bk^2}{\bk^2} \ln \frac{2\rho^+P^-}{\bk^2}\,\nn
 &=& \frac{\bar\alpha_s}{2}\left ( \ln^2\frac{2\rho^+ P^-}{Q_0^2} - \ln^2\frac{2\rho^+ P^-}{Q^2}\right)\,\nn
  &=& \bar\alpha_s\ln \frac{2\rho^+ P^-}{Q_0 Q} \ln \frac{Q^2}{Q_0^2} \,.
 \eeq
which follows from \eqn{eq:S1loop-ms} in momentum space after taking the double logarithmic limit. The $\bk^2$ dependence in the log is the additional term we found earlier as a consequence of the fact that the gluon life time must be shorter than the size of the target. 
As expected the $\rho^+$ dependence cancels out in the sum and we find
\beq
\sigma_{\rm NLO}(x_{\rm Bj},Q^2)& \approx &\bar\alpha_s\left[ \ln\frac{Q^2}{Q_0^2}\ln\frac{q^+}{\rho^+} - \frac{1}{2}\ln^2\frac{Q^2}{Q_0^2}+\ln \frac{2\rho^+ P^-}{Q_0Q} \ln \frac{Q^2}{Q_0^2}\right]\,\nn
&= & \bar\alpha_s \ln\frac{Q^2}{Q_0^2}\ln\frac{2q^+P^-}{Q^2}=\bar\alpha_s\ln\frac{Q^2}{Q_0^2}\ln\frac{1}{x_{\rm Bj}}\,
\eeq
Although simple, this result is quite a satisfactory confirmation of consistency for our approach: it yields the correct double logarithmic structure at NLO and only one collinear log appears at the $\cO(\alpha_s)$ order as required by the RG structure of QCD. In comparison, the ``naive" shockwave approach which omits the collinearly anomalous term in the NLO correction to the dipole operator at small $x$ yields an additional collinear double log: 
\beq
 \sigma_{\rm NLO, naive}(x_{\rm Bj},Q^2) &\approx& \bar\alpha_s\left[\ln\frac{Q^2}{Q_0^2}\ln\frac{q^+}{\rho^+} - \frac{1}{2}\ln^2\frac{Q^2}{Q_0^2}+\ln \frac{\rho^+ P^-}{Q_0^2} \ln \frac{Q^2}{Q_0^2}\right]\,\nn
 &=& \bar \alpha_s\ln\frac{Q^2}{Q_0^2}\ln\frac{1}{x_{\rm Bj}}  + \frac{\bar\alpha_s}{2}\ln^2\frac{Q^2}{Q_0^2}\,,
\eeq
which is of course not the correct result because of the additional double collinear logarithm.

Another way to see how the composite dipole transformation renders the NLO hard factor parametrically small (as it should) is to compute $\bar H_{\rm NLO}$ directly in the double logarithmic approximation. To that aim, one simply has to add to Eq.~\eqref{eq:C1-DL} the counter-term given by $\bar\alpha_s L(\mu)$ resulting from the change of basis as shown in Eq.~\eqref{eq:he-factorization-NLO}, which reads, in the DLA
\begin{align}
    \bar\alpha_s L(\mu)&\approx \bar\alpha_s\int_{Q_0^2}^{Q^2}\frac{\rmd \bk^2}{\bk^2}\ln\left(\frac{\mu^2}{\bk^2}\right)\,\nn
    &=\frac{1}{2}\ln\left(\frac{QQ_0}{\mu^2}\right)\ln\left(\frac{Q^2}{Q_0^2}\right)\,,
\end{align}
such that, combined with Eq.~\eqref{eq:C1-DL}, we find that
\begin{align}
    \bar H_{\rm NLO}(Q^2,\rho^+/q^+,\mu^2)\approx \bar\alpha_s\ln\left(\frac{Q^2}{Q_0^2}\right)\ln\left(\frac{\mu^2}{x_{\rm Bj}\zeta}\right)\,,
\end{align}
Since the factorization scale $\zeta=2\rho^+P^-$ is ultimately chosen of the order of $2q^+P^-$ ($\rho^+\sim q^+$ in a $k^+$ ordered factorization scheme) and $\mu^2\sim Q^2$, we have $\mu^2/\zeta\sim x_{\rm Bj}$ such that $\bar H_{\rm NLO}$ is clearly a small $\mathcal{O}(\alpha_s)$ correction which does not contain neither any high energy logarithm nor any double collinear logarithm (unlike Eq.~\eqref{eq:C1-DL}). In other words, the ``change of basis" we propose eliminates the large negative double logarithms in the NLO impact factor originally given by Eq.~\eqref{eq:C1-DL}.

%%%%%%%%%%%%%%%%%%%%%%%%%%%%%%%%%%%%%%%%%%%%%%%%%%%%%%
\section{Mellin space analysis of the evolution equation for the composite dipole\label{sec:mellin-space}}
%%%%%%%%%%%%%%%%%%%%%%%%%%%%%%%%%%%%%%%%%%%%%%%%%%%%%%

Now that we have checked that the fixed order NLO result yields the correct double logarithmic structure, we want to demonstrate that the NLO high energy evolution of the composite dipole defined by Eq.~\eqref{eq:NLO-BK} (which would appear in the NNLO fixed order calculation) is also free of collinear logarithms. To that aim, it is sufficient to consider the weak scattering limit where the evolution for the composite dipole linearizes in terms of $T\equiv 1-S$. It is also convenient to work in Mellin space ; we denote the Mellin transform of the dipole amplitude and its Mellin inverse as
\beq 
\bar T(\gamma)= \mu_0^2\int \rmd^2 \bx_{12} (\mu_0|\bx_{12}|)^{2(\gamma-1)} (1-\bar S_{12})  \,,\quad \bar T(\bx_{12})=\int_{c-i\infty}^{c+i\infty}\frac{\rmd \gamma}{2\pi i} \bar T(\gamma) (\mu_0 |\bx_{12}|)^{2\gamma}\,,
\eeq
with $\mu_0$ an arbitrary transverse scale introduced to make the Mellin transform dimensionless, which can be thought of as the characteristic scale of the target. 

We first compute from Eq.~\eqref{eq:NLO-BK} the composite $\bar T$ matrix in Mellin space. To do so, we first express the composite dipole amplitude in the case where this amplitude is a pure power, $T(\bx_{12})=(\mu_0 |\bx_{12}|)^{2\gamma}$:
\begin{align}
\bar T(\bx_{12})=T(\bx_{12})-\frac{\alpha_sN_c}{4\pi^2}\int\rmd ^2\bx_3&\frac{\bx^2_{12}}{\bx^2_{13}\bx^2_{32}}  \left( \ln \frac{1}{\bx_{13}^2\bx_{32}^2 \fscal^4}+\frac{\bx^2_{13}-\bx^2_{23}}{\bx^2_{12}} \ln \frac{\bx_{13}^2}{\bx_{32}^2}\right)\nonumber\\
    &\times \left[(\mu_0 |\bx_{12}|)^{2\gamma}-(\mu_0 |\bx_{13}|)^{2\gamma}-(\mu_0|\bx_{32}|)^{2\gamma}\right]\,.\label{eq:composite-Mellin}
\end{align}
The integral over $\bx_3$ can be performed using the generalized result for the conformal ``bubble" integral
\begin{align}
    B(\bx_{12},a,b)\equiv \int\rmd^{D} \bx_3\frac{1}{(\bx_{13})^{2a}(\bx_{32})^{2b}}=\pi^{D/2}\frac{\Gamma(a+b-D/2)\Gamma(D/2-a)\Gamma(D/2-b)}{\Gamma(a)\Gamma(b)\Gamma(D-a-b)}(\bx_{12}^2)^{D/2-a-b}\label{eq:bubble-int}\,,
\end{align}
for arbitrary powers $a$ and $b$. Taking derivative w.r.t.~$a$ or $b$ and setting $a=0$ or $b=0$ brings a logarithm of $\bx_{13}$ or $\bx_{32}$ such that all terms in Eq.~\eqref{eq:composite-Mellin} can be computed analytically in $D$ dimension:
\begin{align}
    &\bar T(\bx_{12})=T(\bx_{12})+\lim\limits_{D\to 2}\left\{-\frac{\alpha_sN_c}{4\pi^2}(\mu_0|\bx_{12}|)^{2\gamma}\left[\left.\bx_{12}^2\left(2\frac{\partial B}{\partial a}-4\ln(\fscal)B\right)\right|_{a=b=1}+2\left.\left(\frac{\partial B}{\partial b}-\frac{\partial B}{\partial a}\right)\right|_{a=0,b=1}\right]\right.\nonumber\\
    &\left.+\frac{\alpha_sN_c}{2\pi^2}\mu_0^{2\gamma}\left[\left.\bx_{12}^2\left(\frac{\partial B}{\partial a}+\frac{\partial B}{\partial b}-4\ln(\mu)B\right)\right|_{a=1,b=1-\gamma}+\left(\left.\frac{\partial B}{\partial b}-\frac{\partial B}{\partial a}\right)\right|_{a=0,b=1-\gamma}-\left.\left(\frac{\partial B}{\partial b}-\frac{\partial B}{\partial a}\right)\right|_{a=1,b=-\gamma}\right]\right\}\,,
\end{align}
where we have used $\bx_1\leftrightarrow\bx_2$ symmetry.
In the end, we find
\beq
\bar T(\bx_{12})=\left[1+\bar\alpha_sf_1(\gamma)-\bar\alpha_s\chi_1(\gamma)\left(\frac{\rmd}{\rmd\gamma }+\ln(\fscal^2/\mu_0^2)\right)\right] (\mu_0 |\bx_{12}|)^{2\gamma}\label{eq:composite-dipole-mellinspace}
\eeq
In this expression, $\chi_1$ is the LO BKFL eigenvalue, defined as
\begin{align}
    \chi_1(\gamma)\equiv 2\psi(1)-\psi(\gamma)-\psi(1-\gamma)\,,
\end{align}
with $\psi(\gamma)$ is the digamma function, $\psi(1)=-\gamma_E$ and $f_1$ is an analytically computable function, given by
\begin{align}
    f_1(\gamma)
    =&-\frac{1}{2\gamma}\left[(2\gamma_E+\pi\cot(\pi\gamma))(2+2\gamma_E\gamma+\pi\gamma\cot(\pi\gamma)+\pi^2\gamma\cot^2(\pi\gamma)\right.\nonumber\\
    &\left.+4\psi(\gamma)(1+2\gamma_E\gamma+\pi\gamma\cot(\pi\gamma)+\gamma\psi(\gamma))-2\gamma\psi'(\gamma)\right]\,,
\end{align}
but whose precise form is actually of no relevance to our analysis. Indeed, the important structure arises from the derivative with respect to $\gamma$ which is tied to the scale dependence (via $\fscal$) of our composite operator. In particular, in the conformal dipole case, the coefficient multiplying the $\rmd/\rmd\gamma$ operator is smaller by a factor of 2 because of the difference in the $\fscal$ dependence between the conformal and our composite dipole. Using Eq.~\eqref{eq:composite-dipole-mellinspace} and taking a derivative with respect to $\ln(\zeta)$, one easily gets the NLO evolution of the composite $\bar T$ matrix in Mellin space since
\beq
  \frac{\partial \bar T (\gamma)}{\partial \ln(\zeta)}=\bar\alpha_s\chi_1(\gamma)\bar T(\gamma)-\bar\alpha_s^2\chi_1(\gamma)\chi'_1(\gamma)\bar T(\gamma)+\bar\alpha_s^2\chi_2(\gamma)\bar T(\gamma)+\mathcal{O}(\alpha_s^3)\,,\label{eq:NLO-BFKL-composite}
\eeq
where we have used that $[\chi_1(\gamma),f_1(\gamma)]=0$ and 
\beq 
\left[\chi_1(\gamma),\chi_1(\gamma)\frac{\rmd}{\rmd\gamma}\right]=-\chi_1(\gamma)\chi'_1(\gamma)\,.
\eeq
In this expression $\chi_2$ is the two-loop $\mathcal{O}(\alpha_s^2)$ BFKL eigenvalue, whose expression can be found in~\cite{Fadin:1998py,Ciafaloni:1998gs,Kotikov:2000pm}. The pathological collinear double logarithm behavior of the NLO BFKL equation manifests itself as a triple pole in $\gamma=1$ in $\chi_2$:
\begin{align}
    \chi_{\rm 2}(\gamma)&=-\frac{1}{(1-\gamma)^3}-\left(\frac{11}{12}+\frac{n_f}{6N_c^3}\right)\frac{1}{(1-\gamma)^2}+\mathcal{O}\left(\frac{1}{1-\gamma}\right)\,.
\end{align}
On the other hand, the additional NLO term introduced by the scale dependence of the composite operator, i.e.~$-\bar\alpha_s^2\chi_1(\gamma)\chi'_1(\gamma)$, also yields a triple pole at $\gamma=1$ which exactly cancels the collinear pole in the NLO BFKL kernel:
\begin{align}
    \chi_1(\gamma)\chi'_1(\gamma)&=-\frac{1}{(1-\gamma)^3}+2\zeta(3)+\mathcal{O}((1-\gamma)^2)\,.
\end{align}
This exact cancellation
relies on the particular value of the coefficient in front of the $\chi_1(\gamma)\chi'_1(\gamma)$ term ; in particular, the NLO evolution of the conformal dipole is not free of double collinear logarithms, which \textit{a posteriori} explains why the numerical solution of the NLO BK equation for the conformal dipole is instable~\cite{Lappi:2015fma}.

However, the NLO evolution equation Eq.~\eqref{eq:NLO-BFKL-composite} contains spurious \textit{anti-collinear} logarithms which manifest in Mellin space as poles when $\gamma\to 0$. The two-loop BFKL eigenvalue behaves likes
\begin{align}
    \chi_2(\gamma)&=-\left(\frac{11}{12}+\frac{n_f}{6N_c^3}\right)\frac{1}{\gamma^2}+\mathcal{O}\left(\frac{1}{\gamma}\right)\,,
\end{align}
while the $\chi_1(\gamma)\chi'_1(\gamma)$ introduces a triple pole in $\gamma=0$
\begin{align}
    \chi_1(\gamma)\chi'_1(\gamma)&=-\frac{1}{\gamma^3}+2\zeta(3)+\mathcal{O}(\gamma^2)\,,
\end{align}
which is responsible, in physical space, to the additional anti-collinear double logarithms in the NLO BFKL equation for the composite dipole. However, such anti-collinear double logarithms are harmless in the non-linear saturation regime as they are suppressed by unitarity.

The main point of this section is that the NLO BFKL equation for the composite dipole in the dilute limit is free from any collinear double logarithms. This property is largely independent of the specific form of the kernel $L_{123}$ used in the basis change between the dipole and the composite dipole, as long as the dependence on the collinear factorization scale $\mu$ remains the same. In the next section, we will leverage this freedom to select an appropriate form for $L_{123}$, aiming to simplify the NLO equation for the composite dipole in the non-linear regime as much as possible.

While a different choice of $L_{123}$ would not affect the NLO BFKL equation for the composite dipole, it leads to a different form for the $f_1(\gamma)$ function which does appear in the NLO impact factor after performing the change of basis between $T$ and $\bar T$. If, instead of Eq.~\eqref{eq:NLO-BK}, one uses a simplified version of the composite dipole with $L_{123}=K_{123}\ln(1/(|\bx_{13}||\bx_{32}|\mu^2)$, then the function $f$ reads
\begin{align}
    f_1(\gamma)=-\frac{1}{4}(4\gamma_E^2+(4\gamma_E+\psi(\gamma)+\psi(1-\gamma))(\psi(\gamma)+\psi(1-\gamma))+3\psi'(1-\gamma)-3\psi'(\gamma))\,.
\end{align}
These findings, together with those of the previous section, can be formalized at the level of the physical NLO DIS cross-section.
According to Eq.~\eqref{eq:composite-dipole-mellinspace}, the $L$ operator (defined by Eq.~\eqref{eq:allorder-composite-dipole}) in Mellin space reads
\beq
 \cM\{L\} = -f_1(\gamma)+\chi_1(\gamma) \left[\frac{\rmd }{\rmd \gamma}+\ln(\mu^2/\mu_0^2)\right]\,,
\eeq
at one loop (i.e.~$L=L_1+\mathcal{O}(\alpha_s)$ in the notations of Sec.~\ref{sec:comp-dip}). After the change of basis, the resulting NLO cross-section in Mellin space becomes then (see also Eq.~\eqref{eq:he-factorization-NLO})
\bea\label{eq:he-factorization-Mellin}
\sigma_{\rm NLO}(x,Q^2)&= \int \frac{\rmd \gamma}{2 \pi i }H_{\rm NLO}(\rho^+/q^+,\gamma)  \, \left[1-\bar\alpha_sf_1(\gamma) +\abar \chi_1(\gamma) \left(\frac{\rmd }{\rmd \gamma}+\ln(\mu^2/\mu_0^2)\right)\right ]\, \nn
&\times \exp\left[\abar (\chi_1(\gamma) -\abar  \chi_1\chi'_1(\gamma)+\abar\chi_2(\gamma))\ln \frac{\rho^+}{\rho^+_0}  \right] \bar T (\gamma, \rho_0^+P^-)+\mathcal{O}(\alpha_s^2)\,.
\eea
where we have used that $\cM\{[K,L] \} = \left[\chi_1(\gamma), \chi_1(\gamma) \frac{\rmd }{\rmd \gamma}\right] =  -\chi_1(\gamma) \chi'_1(\gamma)$.
Integrating by part the leftmost derivative we obtain \bea\label{eq:he-factorization-Mellin-2}
\sigma_{\rm NLO}(x,Q^2)&= \int \frac{\rmd \gamma}{2 \pi i }\left(\bar H_{\rm NLO}(\rho^+/q^+,\gamma)+\mathcal{O}(\alpha_s^2)\right)\exp\left[ \abar\bar\chi_{\rm NLO}(\gamma)\ln \frac{\rho^+}{\rho^+_0}  \right] \bar T (\gamma, \rho_0^+P^-)\,,
\eea
where the new NLO impact factor and NLO eigenvalue respectively read
\begin{align}
    \bar H_{\rm NLO}(\rho^+/q^+,\gamma)&=H_{\rm NLO}(\rho^+/q^+,\gamma)+  H_{\rm LO}(\gamma) \left[\chi_1(\gamma) \ln(\mu^2/\mu_0^2)-f_1(\gamma)\right]-   \frac{\rmd}{\rmd \gamma}\left(\chi_1(\gamma) H_{\rm LO}(\gamma)\right)\,,\\
    \bar\chi_{\rm NLO}(\gamma)&=\chi_{\rm NLO}(\gamma)-\bar\alpha_s\chi_1(\gamma)\chi'_1(\gamma)=\chi_1(\gamma)+\bar\alpha_s\chi_{2}(\gamma)-\bar\alpha_s\chi_1\chi'_1(\gamma)\,.
\end{align}
The NLO impact factor in Mellin space thus receives new contribution coming from the composite dipole transformation, such that the physical cross-section is invariant (up to $\mathcal{O}(\alpha_s^2)$ corrections). In particular, one sees that it is sensitive to the function $f_1(\gamma)$ which depends on the specific choice of kernel $L_{123}$ in the composite dipole (one can also choose $\mu_0=\mu$ in the definition of the Mellin transform to eliminate the $\ln(\mu^2/\mu_0^2)$ term). The derivative operator also brings a NLO correction proportional to the derivative of the the LO impact factor times the LO BFKL eigenvalue.

Finally, the analysis of the collinear structure of the high energy evolution of $T$ carried out in Sec.~\ref{sec:comp-dip} enables one to obtain the NNLO evolution equation for the composite $\bar T$ matrix, given the three-loop BFKL eigenvalue $\chi_3(\gamma)$ which has been computed in \cite{Gromov:2015vua,Caron-Huot:2016tzz} in $\mathcal{N}=4$ supersymmetric Yang-Mills Theory (see also~\cite{Marzani:2007gk} for an approximate expression for QCD and note that the linearization of the results obtained in~\cite{brunello2025highenergyevolutionplanarqcd}, together with the conformal and matter loops contributions, should provide the exact analytic result in QCD). Following Eq.~\eqref{eq:L2-formal}, we define the $\mathcal{O}(\alpha_s^2)$ contribution to the $L$ operator in Mellin space as
\begin{align}
   \mathcal{M}\{L_2\}=-f_2(\gamma)+\left\{\chi_2(\gamma)-\frac{1}{2}\chi_1(\gamma)\chi'_1(\gamma)\right\}\left[\frac{\rmd}{\rmd \gamma}+\ln(\mu^2/\mu_0^2)\right]\,,
\end{align}
where the function $f_2$ is scheme-dependent and does not matter for the NNLO BFKL kernel, while the $\ln(\mu^2)$ dependence (and thus, the factor in front of the $\rmd/\rmd \gamma$ operator) is constrained to be the operator $K_2+\frac{1}{2}[K_1,L_{10}]$ in Mellin space, as demonstrated in Sec.~\ref{sec:comp-dip}. Combining this relation with Eq.~\eqref{eq:KNNLO-formal}, we find that the NNLO BFKL eigenvalue for the composite $\bar T$ matrix reads
\begin{align}
    \bar\chi_{\rm NNLO}(\gamma)=\ &\chi_{\rm NNLO}(\gamma)-\bar\alpha_s\chi_1(\gamma)\chi'_1(\gamma)-\bar\alpha_s^2 \left[\chi_2(\gamma)\chi'_1(\gamma)+\chi'_2(\gamma)\chi_1(\gamma)- \chi_1(\gamma)\chi_1'^2(\gamma)-\frac{1}{2} \chi_1^2(\gamma) \chi_1''(\gamma)\right]\,,\label{eq:NNLO-BFKL-safekernel}
\end{align}
with $\chi_{\rm NNLO}(\gamma)=\chi_1(\gamma)+\bar\alpha_s\chi_2(\gamma)+\bar\alpha_s^2\chi_3(\gamma)$. Eq.~\eqref{eq:NNLO-BFKL-safekernel} does not depend on $f_2$ nor $\ln\mu^2$ and is free of collinear poles associated with double collinear logarithms by construction. Eq.~\eqref{eq:NNLO-BFKL-safekernel} coincides with Eq.~(46) of \cite{Deak:2019wms} upon the replacement $\rmd/\rmd\gamma \to \frac{1}{2}\rmd/\rmd\gamma$, thereby confirming that our framework naturally incorporates the scale redefinition proposed in the BFKL context. The factor of 2 in front of the $\gamma$-derivative arises because our approach effectively corresponds to a change of variables from one asymmetric choice ($k^+$-ordered) to another asymmetric choice ($k^-$-ordered), rather than from a symmetric to an asymmetric choice, as considered in \cite{Deak:2019wms}.

%%%%%%%%%%%%%%%%%%%%%%%%%%%%%%%%%%%%%%%%%%%%%%%%%%%%%%
\section{NLO BK equation for the composite dipole\label{sec:sudakov}}
%%%%%%%%%%%%%%%%%%%%%%%%%%%%%%%%%%%%%%%%%%%%%%%%%%%%%%

In this section, we present the NLO BK equation for the composite dipole which becomes free of double collinear logarithms. From a formal standpoint, as discussed in Sec.~\ref{sec:he-factorization}, we present here the calculation of the commutator between the LO BK kernel $K_{123}$ and the kernel appearing in the composite dipole denoted as $L_{123}$. Unlike in the dilute BKFL regime though, the precise form of the composite dipole matters in the non-linear regime. For this reason, we explore the dependence of the NLO BK equation on the choice of the transverse scale $\br_{123}$ accompanying the arbitrary scale $\mu$ in the definition of the composite dipole. This general choice of composite dipole is given by Eq.~\eqref{eq:compositedipole-general}, which we rewrite here as
\begin{align}
    \bar S_{12}&= S_{12} +  \abar \int_3 K_{123} \ln\left(\frac{1}{\br_{123}^2\mu^2}\right) \left[ S_{13}S_{32}-S_{12}\right]\,,
\end{align}
with $\br_{123}$ some transverse coordinate scale defined in terms of $\bx_{1},\bx_{2}$ and $\bx_3$. Taking a derivative with respect to $\ln(\zeta)$ and working at NLO (we neglect $\mathcal{O}(\alpha_s^3)$ terms), the NLO BK equation for $\bar S_{12}$ reads
\begin{align}
    \frac{\partial \bar S_{12}}{\partial \ln(\zeta)}&=\abar K_{\rm NLO}\otimes\bar S_{12}+\abar^2\int_{3,4} K_{123}K_{134}\ln\left(\frac{\br_{134}^2}{\br_{123}^2}\right)(\bar S_{14}\bar S_{43}-\bar S_{13})\bar S_{32}\nonumber\\
    &+\abar^2\int_{3,4} K_{123}K_{324}\ln\left(\frac{\br_{324}^2}{\br_{123}^2}\right)(\bar S_{34}\bar S_{42}-\bar S_{32})\bar S_{13}-\abar^2\int_{3,4} K_{123}K_{124}\ln\left(\frac{\br_{124}^2}{\br_{123}^2}\right)(\bar S_{14}\bar S_{42}-\bar S_{12})\,.\label{eq:NLO-BK-comp-step1}
\end{align}
In this equality the second term (and similarly third and fourth) come from the replacement of $S_{13}$ (resp. $S_{32}$ and $S_{12}$) in the LO BK equation by its expression in terms of $\bar S_{13}$ (resp. $\bar S_{32}$ and $\bar S_{12}$) and the $\ln(\zeta)$ derivative of $S_{13}$ (resp. $S_{32}$ and $S_{12}$) present in the definition of the composite dipole as given by the LO BK equation. To the accuracy of interest, one can then replace all the ``bare" dipoles by the composite ones in the terms of order $\alpha_s^2$, in particular for the term given by the NLO BK kernel. After these manipulations, the NLO BK equation for the composite dipole is manifestly $\mu$ independent.

While Eq.~\eqref{eq:NLO-BK-comp-step1} is general, it is not the most convenient form because each term contains UV divergences which only cancel in the sum. In order to obtain a NLO BK equation which is manifestly free of UV divergences term by term, we wish to reorganize Eq.~\eqref{eq:NLO-BK-comp-step1} following the color structures present in the conventional NLO BK equation written in Eqs.~\eqref{eq:NLO-BK}. To do so, we first isolate the two terms depending on the product of three dipole, $\bar S_{34}\bar S_{42}\bar S_{13}$ and $\bar S_{14}\bar S_{43}\bar S_{32}$. By exchanging $\br_3\leftrightarrow \br_4$ in the latter contribution, one realizes that these two triple dipole term can be combined together ; this yields a kernel of the form $K_{123}K_{324}\ln[\br_{324}^2\br_{143}^2/(\br_{123}^2\br_{124}^2)]$. In Eq.~\eqref{eq:NLO-BK}, the color structure depending on the product of three dipoles is actually $(\bar S_{34}\bar S_{42}-\bar S_{32})\bar S_{13}$ ; therefore, we further add and subtract a double dipole $\bar S_{13}\bar S_{32}$ with the same kernel. Finally, we group together all the remaining terms depending on the double dipoles. In the end, we find that the previous equation can be re-organized according to
\begin{align}
    \frac{\partial \bar S_{12}}{\partial \ln(\zeta)} &=\abar K_{\rm NLO}\otimes\bar S_{12}+\bar\alpha_s^2\int_{3,4} K_{123}K_{324}\ln\left(\frac{\br_{324}^2\br_{143}^2}{\br_{123}^2\br_{124}^2}\right)(\bar S_{34}\bar S_{42}-\bar S_{32})\bar S_{13}\nonumber\\
    &-\bar\alpha_s^2\int_{3,4} K_{123}\left[K_{134}\ln\left(\frac{\br_{134}^2}{\br_{123}^2}\right)+ K_{324}\ln\left(\frac{\br_{124}^2}{\br_{143}^2}\right)+K_{124}\ln\left(\frac{\br_{123}^2}{\br_{124}^2}\right)\right](\bar S_{13}\bar S_{32}-\bar S_{12})\,.\label{eq:NLO-BK-comp}
\end{align}
Note that, by anti-symmetry under $\bx_3\leftrightarrow \bx_4$ exchange, we have
\begin{align}
    \int_{3,4} K_{123}\left[K_{134}\ln\left(\frac{\br_{134}^2}{\br_{123}^2}\right)+ K_{324}\ln\left(\frac{\br_{124}^2}{\br_{143}^2}\right)+K_{124}\ln\left(\frac{\br_{123}^2}{\br_{124}^2}\right)\right]=0\,,
\end{align}
such that one can  add for free the single dipole term $\bar S_{12}$ in the subtraction term of the double dipole contribution depending on $\bar S_{13}\bar S_{32}$. 

We now discuss the choices of $\br_{123}$ such that (i) the integral of $\br_{4}$ in the third term depending on the color structure $\bar S_{13}\bar S_{32}-\bar S_{12}$ can be performed analytically, (ii) each term in Eq.~\eqref{eq:NLO-BK-comp} is separately UV finite. All the forms of $\br_{123}$ considered in this section lead to a NLL BK equation for the composite dipole which is free of collinear double logarithms ; yet this is not necessarily manifest term by term in the evolution equation since depending on the choice of $\br_{123}$, double collinear logarithms can remain both in the triple dipole term depending on $\bar S_{13}(\bar S_{34}\bar S_{42}-\bar S_{32})$ and in the double dipole one, such that they only cancel in the sum.

%Now, there are choices of the transverse vector $\br_{123}$ such that the integral over $\bx_4$ in the second line of Eq.~\eqref{eq:NLO-BK-comp} can be performed analytically. 
For instance, if one uses a value for $\br_{123}$ inspired by the conformal dipole considered in~\cite{Balitsky:2009xg}:
\begin{align}
    \br_{123}^2=\frac{\bx_{13}^2\bx_{32}^2}{\bx_{12}^2}\,,\label{eq:BC-r123-choice}
\end{align}
then, one can use the following integral computed in \cite{Balitsky:2009xg}
\begin{align}
   \int_4\left[\frac{\bx_{13}^2}{\bx_{14}^2\bx_{34}^2}\ln\left(\frac{\bx_{14}^2\bx_{34}^2\bx_{12}^2}{\bx_{13}^4\bx_{32}^2}\right)+ \frac{\bx_{32}^2}{\bx_{34}^2\bx_{24}^2}\ln\left(\frac{\bx_{14}^4\bx_{24}^2}{\bx_{13}^2\bx_{34}^2\bx_{12}^2}\right)+\frac{\bx_{12}^2}{\bx_{14}^2\bx_{24}^2}\ln\left(\frac{\bx_{13}^2\bx_{32}^2}{\bx_{14}^2\bx_{42}^2}\right)\right]=& \ -\ln\left(\frac{\bx_{12}^2}{\bx_{13}^2}\right)\ln\left(\frac{\bx_{12}^2}{\bx_{32}^2}\right)\nonumber\\
   &-4\zeta(3)\delta(\bx_{13})\,.\label{eq:BC-identity}
\end{align}
The term proportional to $\delta(\bx_{13})$ does not contribute to the remaining integral over $\br_3$ thanks to unitarity. Half of the double log on the right hand side of Eq.~\eqref{eq:BC-identity} is canceled by the double log present in the $K_{\rm NLO}$ term, which reads (recall Eq.~\eqref{eq:K123-NLO-def})
\begin{align}
    K_{\rm NLO}\otimes\bar S_{12} =  \int_3 K_{123}\left[1-\frac{\abar}{2}\ln\left(\frac{\bx_{12}^2}{\bx_{13}^2}\right)\ln\left(\frac{\bx_{12}^2}{\bx_{32}^2}\right)\right](\bar S_{13}\bar S_{32}-\bar S_{12})+...\label{eq:KNLO-DL}
\end{align}
while the other half gets canceled by the second term in Eq.~\eqref{eq:NLO-BK-comp} depending on the triple dipole color structure (this cancellation would be manifest in the linear regime, as shown by the Mellin space analysis).

That being said, the simplest choice for $\br_{123}$ would be the parent dipole size, namely,
\begin{align}
    \br_{123}^2=\bx_{12}^2\,.
\end{align}
In this case, in order to simplify the term in Eq.~\eqref{eq:NLO-BK-comp} depending on the color structure $\bar S_{13}\bar S_{32}-\bar S_{12}$, one must evaluate the integral
\begin{align}
    \frac{1}{\pi}\int\rmd^2\bx_4\left[\frac{\bx_{13}^2}{\bx_{14}^2\bx_{34}^2}\ln\left(\frac{\bx_{13}^2}{\bx_{12}^2}\right)+ \frac{\bx_{32}^2}{\bx_{34}^2\bx_{24}^2}\ln\left(\frac{\bx_{12}^2}{\bx_{14}^2}\right)\right]\,,
\end{align}
which is divergent in the UV as the logarithmic singularity in the limit $\bx_{4}\to \bx_{1}$ of the first term does not cancel with the second term. This UV divergence gets canceled by a similar divergence in the term depending on the color structure $(\bar S_{34}\bar S_{42}-\bar S_{32})\bar S_{13}$ ; yet, this demonstrates that not all $\br_{123}$ scales are convenient from a numerical point of view, since it would be preferable that each term in Eq.~\eqref{eq:NLO-BK-comp} is separately UV and IR finite.

A sufficient condition on $\br_{123}$ for the integral over $\bx_4$ to be UV finite is that it satisfies
\begin{align}
  \forall \bx_1,\bx_2,\bx_3\,,\quad   \br_{134}^2\underset{\bx_4\to \bx_1}{\sim}\br_{124}^2\,,\label{eq:r123-condition}
\end{align}
which is indeed true for the choice given by Eq.~\eqref{eq:BC-r123-choice} but obviously not for $\br_{123}^2=\bx_{12}^2$. Importantly, this condition also ensures that the second term in Eq.~\eqref{eq:NLO-BK-comp} is UV finite for $\bx_3\to \bx_1$ (note that the other singular points $\bx_4\to \bx_3$ and $\bx_4\to \bx_2$ of the kernel $K_{123}K_{324}$ are cured by unitarity).

Possible expressions for $\br_{123}$ satisfying this condition are $\br_{123}^2=\min(\bx_{13}^2,\bx_{23}^2)$, the choice given by Eq.~\eqref{eq:NLO-BK} emerging from the $x\to 0$ limit of the $x$-dependent dipole distribution computed in section~\ref{sec:he-factorization}, which reads
\beq
\br_{123}^2=|\bx_{13}||\bx_{32}|\exp\left[\frac{\bx_{23}^2-\bx_{13}^2}{2\bx_{12}^2}\ln\left(\frac{\bx_{13}^2}{\bx_{23}^2}\right)\right]\,,
\eeq
or the generalization of Eq.~\eqref{eq:BC-r123-choice} with an arbitrary power $\beta$
\begin{align}
    \br_{123}^2=\bx_{13}^2\left(\frac{\bx_{32}^2}{\bx_{12}^2}\right)^\beta\,.\label{eq:r123-classbeta}
\end{align}
In the two former cases, the drawback is that we did not manage to find a closed analytic expression for the $\bx_4$ integral in the double dipole term. Among the latter class given by Eq.~\,\eqref{eq:r123-classbeta}, there are two interesting cases. The first one is $\beta=-1$ since then, one can also easily perform the $\bx_4$ integral using the standard result for the conformal bubble integral in $d$ dimension:
\begin{align}
    \int_4\left[\frac{\bx_{13}^2}{\bx_{14}^2\bx_{34}^2}\ln\left(\frac{\bx_{14}^2\bx_{32}^2}{\bx_{12}^2\bx_{34}^2}\right)+ \frac{\bx_{32}^2}{\bx_{34}^2\bx_{24}^2}\ln\left(\frac{\bx_{12}^2\bx_{34}^2}{\bx_{24}^2\bx_{13}^2}\right)+\frac{\bx_{12}^2}{\bx_{14}^2\bx_{24}^2}\ln\left(\frac{\bx_{13}^2\bx_{24}^2}{\bx_{14}^2\bx_{23}^2}\right)\right]=0\,.
\end{align}
For this choice, the additional term depending on the color structure $\bar S_{13}\bar S_{32}-\bar S_{12}$ simply vanishes, and the double collinear log in the color structure $\bar S_{13}\bar S_{32}-\bar S_{12}$ of the NLO BK kernel remains as it stands, but gets canceled by a doube collinear logarithm ``hidden" in the triple dipole term of the NLO evolution of the composite dipole.

For $\beta=0$, we have $\br_{123}^2=\bx_{13}^2$ and the integral of $\bx_4$ one must compute involves not only bubble graphs but also a ``triangle" graph with three UV singular points at $\bx_4=\bx_3,\bx_2$ and $\bx_1$ from the second term of this expression:
\begin{align}
\int_4\left[\frac{\bx_{13}^2}{\bx_{14}^2\bx_{34}^2}\ln\left(\frac{\bx_{14}^2}{\bx_{13}^2}\right)+ \frac{\bx_{32}^2}{\bx_{34}^2\bx_{24}^2}\ln\left(\frac{\bx_{14}^2}{\bx_{13}^2}\right)+\frac{\bx_{12}^2}{\bx_{14}^2\bx_{24}^2}\ln\left(\frac{\bx_{13}^2}{\bx_{14}^2}\right)\right]=-\frac{1}{2}\ln\left(\frac{\bx_{12}^2}{\bx_{13}^2}\right)\ln\left(\frac{\bx_{12}^2}{\bx_{23}^2}\right)\,.\label{eq:beta=0-x4int}
\end{align}
Note the difference of factor of 2 with Eq.~\eqref{eq:BC-identity}, corresponding to $\beta=1$. The integral Eq.\,\eqref{eq:beta=0-x4int} is computed in appendix~\ref{app:integrals}.

Therefore, interestingly, the case $\beta=0$, i.e.~$\br_{123}^2=\bx_{13}^2$, enables one to exactly cancel the double log contribution in the NLO BK kernel \textit{color structure per color structure}. This choice of the composite dipole should be favored for numerical implementation of the NLO BK equation. With this choice the structure $(\bar S_{34}\bar S_{42}-\bar S_{32})\bar S_{13}$ in the NLO BK equation for the composite dipole is also quite simple and does not contain any double collinear logarithm as compared to the other values of $\beta$ previously discussed ; it is given by
\begin{align}
    \bar\alpha_s^2\int_{3,4} K_{123}K_{324}\ln\left(\frac{\bx_{34}^2}{\bx_{14}^2}\right)(\bar S_{34}\bar S_{42}-\bar S_{32})\bar S_{13}\,.
\end{align}
and is also clearly UV finite.

The only drawback of Eq.~\eqref{eq:r123-classbeta} is that it breaks the $\bx_1\leftrightarrow \bx_2$ symmetry when considering the evolution equation for the real part of composite dipole, i.e.~neglecting the Odderon component~\cite{Hatta:2005as}. This symmetry can easily be restored by taking $r_{123}^2=|\bx_{13}| |\bx_{32}|$, or equivalently by (i) symmetrizing the definition of the composite dipole when $\br_{123}$ is given by Eq.~\eqref{eq:r123-classbeta} and $\beta=0$, namely,
\begin{align}    
\bar S_{12}=S_{12}+\frac{\abar}{2} \int_3 K_{123} \ln\left(\frac{1}{\bx_{13}^2\bx_{32}^2\mu^4}\right) \left[ S_{13}S_{32}-S_{12}\right]    \,,\label{eq:composite-dipole-numeric}
\end{align}
and (ii) symmetrizing the manipulations between Eq.~\eqref{eq:NLO-BK-comp-step1} and Eq.~\eqref{eq:NLO-BK-comp}. In the end, the NLO BK equation for the composite dipole Eq.~\eqref{eq:composite-dipole-numeric} is
\begin{align}\label{eq:comp-evol}
    \frac{\partial \bar S_{12}}{\partial \ln(\zeta)}&=\abar \tilde K_{\rm NLO}\otimes\bar S_{12}+\frac{\bar\alpha_s^2}{2}\int_{3,4} K_{123}K_{324}\ln\left(\frac{\bx_{34}^2}{\bx_{14}^2}\right)\bar S_{13}(\bar S_{34}\bar S_{42}-\bar S_{32})\nonumber\\
    &+\frac{\bar\alpha_s^2}{2}\int_{3,4} K_{123}K_{134}\ln\left(\frac{\bx_{34}^2}{\bx_{24}^2}\right)(\bar S_{14}\bar S_{43}-\bar S_{13})\bar S_{32}\,,
\end{align}
where $\tilde K_{\rm NLO}$ is the NLO BK kernel \textit{without} the double collinear logarithm in Eq.~\eqref{eq:KNLO-DL}. 

Although the composite dipole evolution is now free from collinear logarithms, the transformation in Eq.~\eqref{eq:kernel-transformation} introduces anti-collinear double logarithms, as observed in the BFKL regime in Mellin space. These are suppressed by non-linear dynamics and are therefore innocuous when saturation physics is included.

%%%%%%%%%%%%%%%%%%%%%%%%%%%%%%%%%%%%%%%%%%%%
\section{Conclusion and Outlook}
%%%%%%%%%%%%%%%%%%%%%%%%%%%%%%%%%%%%%%%%%%%%
In summary, we have developed a systematic operator-based framework for high-energy QCD factorization in dilute--dense collisions that eliminates the pathological collinear double logarithms responsible for numerical instabilities and negative cross-sections at NLO. By redefining the dipole operator through a covariant transformation, we construct a composite dipole that restores time ordering and guarantees scheme-independent physical cross sections to all orders in perturbation theory.  

Previous approaches to this problem relied on \textit{ad hoc} prescriptions, such as direct resummation of logarithms, the imposition of kinematic constraints in the Balitsky--Kovchegov (BK) equation, or modifications of the rapidity-evolution variable. While partially effective, these methods depart from the systematic structure of the high-energy OPE and lack a transparent path toward generalization at higher orders. Moreover, they concentrate narrowly on BK evolution without addressing the underlying factorization scheme. A central insight of our framework is that the evolution of hadronic operators and the definition of the form factor are inseparably linked: modifying one necessarily requires a consistent redefinition of the other to preserve the invariance of physical cross sections. By introducing a covariant transformation of CGC operators and their form factors within the OPE, we provide a systematic method to track collinear logarithmic terms order by order in perturbation theory.  

Before presenting the general systematics of our approach in Sec.~\ref{sec:results}, we revisited the high-energy limit that gave rise to the CGC framework, where operators are constructed from infinite Wilson-line correlators. To this end, we introduced a generalized dipole operator that explicitly retains its dependence on $x$ --- the conjugate variable to the lifetime of gluon fluctuations in the target --- but reduces to the standard CGC dipole operator $\langle \mathrm{Tr}(U_{\bx_1} U^\dagger_{\bx_2}) \rangle$ in the limit $x \to 0$. Preserving this explicit $x$ dependence effectively accounts for the finite size of the shockwave, thereby restoring the time-ordering constraint absent in the conventional CGC setup, where the shockwave is assumed to be infinitely thin.  

Analyzing the one-loop correction to the generalized dipole operator, we found that, in addition to the standard LO BK term, an extra contribution arises involving a transverse logarithm, $\ln(\bk^2/\mu^2)$, of the ratio between the transverse momentum of the radiated gluon and the arbitrary collinear factorization scale $\mu$. This logarithm persists in the $x \to 0$ limit and reflects the phase-space restriction on gluon radiation imposed by time ordering. This instructive result motivates the more general operator-based framework developed in this work.  

Our central proposal is to perform a rotation in the space of CGC operators that transforms the naive dipole into a ``composite dipole,'' whose NLO expansion consists of the naive dipole (which obeys the BK equation) supplemented by the additional transverse-logarithm contribution identified above. In the limit $x \to 0$, this composite dipole becomes equivalent to the generalized dipole, not only at LO but also at one loop (up to non-logarithmically enhanced $\mathcal{O}(\alpha_s)$ terms). Importantly, the composite dipole retains the structural simplicity of CGC operators defined at $x=0$: it is expressed solely in terms of infinite Wilson lines in the background field of the shockwave target, a feature that greatly simplifies higher-loop computations in the CGC.  

This construction can be interpreted as a change of basis in the space of CGC operators. By design, physical observables remain invariant under such transformations. In this new basis, both the impact factor and the high-energy renormalization group (RG) kernel acquire NLO corrections that cancel the large double collinear logarithms present in the original formulation. Effectively, the transverse logarithm is reshuffled into the form factor and the initial condition of the RG equation, which thus becomes $\mu$-dependent. This mechanism is reminiscent of the QCD collinear anomaly introduced in~\cite{Becher:2009cu,Gardi:2009qi} and discussed in~\cite{Caron-Huot:2015bja} in the context of non-global logarithms resummation.

We define the composite dipole to all orders as
\beq
\bar S_{12}(\zeta,\mu^2) = \exp\!\left[-\abar L(\abar,\mu^2)\right] S_{12}(\zeta)\,,
\eeq
where $L(\abar,\mu^2)$ is a scale-dependent functional operator in the space of CGC operators, related to the JIMWLK Hamiltonian. Its action can be interpreted as a factorization scheme choice and is constrained by requiring the scaling 
\beq
\bar S_{12}(\zeta,\mu) =  \bar S_{12}\!\left(\frac{\zeta}{\mu^2}\right),
\eeq
with $\zeta=\rho^+ P^-$, where $\rho^+$ plays the role of a factorization scale associated with the $+$-momentum component. This $P^-$-dependent scaling effectively corresponds to changing the ordering variable from $k^+$ to $k^-$, a choice often referred to in the literature as the DIS scheme. As a consequence of the above scaling, the evolution kernel must be $\mu$-independent, i.e. $\bar K(\abar,\mu^2)\to\bar K(\abar)$.  

Concerning the resolution of the collinear instability, the precise form of the composite dipole is not unique and reflects a choice of subtraction scheme. What ultimately matters is the dependence of the composite dipole on the collinear factorization scale $\mu$. We have considered several variants, all sharing the same $\ln(\mu^2)$-dependence but differing by finite $\mathcal{O}(\alpha_s)$ corrections, and computed their associated NLO high-energy RG equations in the nonlinear regime. By construction, these equations do not exhibit large double collinear logarithms and are therefore suitable for numerical evaluation. A particularly simple choice of composite dipole and RG equation is given by Eqs.~\eqref{eq:composite-dipole} and \eqref{eq:composite-dipole-RG}.  As a main application, we derived a new NLO BK equation \eqref{eq:composite-dipole-RG} together with the associated transformed form factor for DIS cross-sections, Eq.~\eqref{eq:barHNLO}. Their numerical study and comparison to data is left for future work.  

Additional outlooks include deriving the collinear-safe evolution equation at NNLO by translating the recently obtained results into our factorization scheme~\cite{brunello2025highenergyevolutionplanarqcd}, as well as extending the framework to other observables such as forward hadron production in proton--nucleus collisions and forward dijet production in electron--proton/nucleus scattering. We anticipate that this framework will provide the foundation for controlled higher-order calculations of hadronic observables at small~$x$, particularly in view of the upcoming Electron--Ion Collider (EIC).

%%%%%%%%%%%%%%%%%%%%%%%%%%%%%%%%%%%%%%%%%%%%
\section*{Acknowledgements}
%%%%%%%%%%%%%%%%%%%%%%%%%%%%%%%%%%%%%%%%%%%%

We thank Guillaume Beuf, Giovanni Chirilli, Piotr Korcyl, Farid Salazar and Anna Sta\`sto for valuable discussions related to this work. Y.~M.~T. was supported by the U.S. Department of Energy under Contract No. DE-SC0012704 and by Laboratory Directed Research and Development (LDRD) funds from Brookhaven Science Associates. We are grateful for the support of the Saturated Glue (SURGE) Topical Theory Collaboration, funded by the U.S. Department of Energy, Office of Science, Office of Nuclear Physics. P.~C. thank the EIC theory institute at BNL
for its support during the initial stages of this work and the Aspen Center for Physics, which is supported by National Science Foundation grant PHY-2210452, where this work was completed.
%%%%%%%%%%%%%%%%%%%%%%
\appendix

%%%%%%%%%%%%%%%%%%%%%%%%%%%%%%%%%%%%%%%%%%%%%%%%
\section{The DIS impact factor at NLO}\label{app:HNLO}
%%%%%%%%%%%%%%%%%%%%%%%%%%%%%%%%%%%%%%%%%%%%%%

In this appendix we give the explicit expression for the NLO correction to the longitudinal DIS cross-section, which can be found in~\cite{Beuf:2017bpd,Hanninen:2017ddy,Caucal:2024cdq},
\begin{align}
   &\Delta \sigma_{L}(x,Q^2)=\Delta H_{\rm NLO}(Q^2,\zeta/s)\otimes S_{\rm LL}(\zeta)=\sum_{i}\frac{4\alpha_{\rm em}e_i^2 N_cS_\perp}{\pi^2}\int\rmd^2\bx_{12}\int_0^1\rmd z_1 \ z_1(1-z_1)\bar Q^2K_0^2(\bar Qx_{12})(1-S_{12})\nonumber\\
   &\times\frac{\alpha_sC_F}{2\pi}\left[\ln^2\left(\frac{z_1}{1-z_1}\right)+5-\frac{\pi^2}{3}\right]+\sum_i\frac{4\alpha_{\rm em}e_i^2N_cS_\perp}{\pi^2 } \int\rmd^2\bx_{12}\int_0^1\rmd z_1\frac{\alpha_sC_F}{\pi^2}\int_0^{1-z_1}\frac{\rmd z_g}{(z_g)_{+,\zeta/s}}\int\rmd^2\bx_3  \ z_1^2z_2^2Q^2\nonumber\\   
     &\times\left\{\left(1+\frac{z_g}{z_1}+\frac{z_g^2}{2z_1^2} \right) \frac{1}{\bx_{13}^2}  \left[K_0^2(QX_R)\left(1-\frac{N_c}{2C_F}S_{13}S_{32}+\frac{S_{12}}{2N_cC_F}\right)\right.-K_0^2(\bar Q_{\rm R}x_{12})e^{-\frac{\bx_{13}^2}{e^{\gamma_E}{\bx_{12}^2}}}(1-S_{12})\right]\nonumber\\
     &+\left(1+\frac{z_g}{z_2}+\frac{z_g^2}{2z_2^2}\right)\frac{1}{\bx_{32}^2}  \left[K_0^2(QX_R) \left(1-\frac{N_c}{2C_F}S_{13}S_{32}+\frac{S_{12}}{2N_cC_F}\right)-K_0^2(\bar Q x_{12})e^{-\frac{\bx_{32}^2}{e^{\gamma_E}{\bx_{12}^2}}}(1-S_{12})\right]\nonumber\\
     &\left.-2\left(1+\frac{z_g}{2z_1}+\frac{z_g}{2z_2} \right) \frac{\bx_{13}\cdot\bx_{32}}{\bx_{13}^2\bx_{32}^2}K_0^2(QX_R) \left(1-\frac{N_c}{2C_F}S_{13}S_{32}+\frac{S_{12}}{2N_cC_F}\right)\right\}\,,\label{eq:HNLO-sigmaL}
\end{align}
where we use the notation
\begin{align}
    \int_0^a\frac{\rmd z_g}{(z_g)_{+,z}}f(z_g)&\equiv  \int_0^a\frac{\rmd z_g}{z_g}\left[f(z_g)-f(0)\Theta(z-z_g)\right]\,,
\end{align}
for any function $f$ of the NLO gluon longitudinal momentum fraction $z_g=k^+/q^+$ which is regular as $z_g\to 0$. In this expression, we use the shorthand notations $z_2=1-z_1-z_g$ and $X_R$ for the effective size of the $q\bar q g$ dipole, which reads $X_R^2=z_1z_2\bx_{12}^2+z_1z_g \bx_{13}^2+z_2z_g\bx_{32}^2$. We also note $\bar Q_R^2=z_1z_2Q^2$ (which is not identical to $\bar Q^2=z_1(1-z_1)Q^2$ except when $z_g=0$ since $z_2+z_g=1-z_1$). It is easy to check that Eq.~\eqref{eq:HNLO-sigmaL} is free of any UV nor IR divergence such that it is suitable for direct numerical evaluation. In particular, there is no divergence as $z_g\to 0$ thanks to the $1/(z_g)_+$ prescription.

Clearly, $\sigma_L(x,Q^2)$ given by Eq.\,\eqref{eq:he-factorization-nlo} must not depend on $\zeta$. Taking a partial derivative of the right-hand side and using the general result
\begin{align}
    \frac{\partial }{\partial\ln(\zeta)}\left[\int_0^a\frac{\rmd z_g}{(z_g)_{+,\zeta/s}}f(z_g)\right]=-f(0)\Theta(a-\zeta/s)\,,
\end{align}
yields the LL renormalization group equation of $S(\bx_{12},\zeta)$ given by Eq.~\eqref{eq:lo-bk}.

%%%%%%%%%%%%%%%%%%%%%%%%%%%%%%%%%%%%%%%%
\section{$x$-dependent gluon distribution at one loop\label{sec:evolutionequation} }
%%%%%%%%%%%%%%%%%%%%%%%%%%%%%%%%%%%%%%%%
Let us consider the $x$-dependent gluon distribution in coordinate
space: 
\begin{align}
S(x,\boldsymbol{x}_{1},\boldsymbol{x}_{2}) & \equiv\frac{1}{N_{c}}\int{\rm d}z_{1}^{+}\int_{-\infty}^{z_{1}^{+}}\!{\rm d}z_{2}^{+}\,{\rm e}^{ixP^{-}z_{12}^{+}}\frac{\partial^{2}}{\partial z_{1}^{+}\partial z_{2}^{+}}{\rm tr}[z_{1}^{+},z_{2}^{+}]_{\boldsymbol{x}_{1}}[z_{2}^{+},z_{1}^{+}]_{\boldsymbol{x}_{2}}\,.\label{eq:dipole-def}
\end{align}
We will use the semi-classical background field methods to derive
its evolution equation. Let us thus decompose the gluon field into
its classical part $A_{{\rm cl}}$ and its quantum part $a$:
\begin{equation}
A^{\mu}=A_{{\rm cl}}^{\mu}+a^{\mu}\,,\label{eq:gluon-field-separation}
\end{equation}
and expand the lines in powers of the quantum field $a$
\begin{align}
[z_{1}^{+},z_{2}^{+}]_{\boldsymbol{x}_{1}} & =[z_{1}^{+},z_{2}^{+}]_{\boldsymbol{x}_{1}}^{{\rm cl}}+ig\int_{z_{2}^{+}}^{z_{1}^{+}}\!{\rm d}z_{3}^{+}[z_{1}^{+},z_{3}^{+}]_{\boldsymbol{x}_{1}}^{{\rm cl}}a^{-}(z_{3}^{+},\boldsymbol{x}_{1})[z_{3}^{+},z_{2}^{+}]_{\boldsymbol{x}_{1}}^{{\rm cl}}\label{eq:expanded-line-1}\\
 & -g^{2}\int_{z_{2}^{+}}^{z_{1}^{+}}\!{\rm d}z_{3}^{+}\int_{z_{2}^{+}}^{z_{3}^{+}}\!{\rm d}z_{4}^{+}[z_{1}^{+},z_{3}^{+}]_{\boldsymbol{x}_{1}}^{{\rm cl}}a^{-}(z_{3}^{+},\boldsymbol{x}_{1})[z_{3}^{+},z_{4}^{+}]_{\boldsymbol{x}_{1}}^{{\rm cl}}a^{-}(z_{4}^{+},\boldsymbol{x}_{1})[z_{4}^{+},z_{2}^{+}]_{\boldsymbol{x}_{1}}^{{\rm cl}}\,,\nonumber 
\end{align}
and
\begin{align}
[z_{2}^{+},z_{1}^{+}]_{\boldsymbol{x}_{2}} & =[z_{2}^{+},z_{1}^{+}]_{\boldsymbol{x}_{2}}^{{\rm cl}}-ig\int_{z_{2}^{+}}^{z_{1}^{+}}\!{\rm d}z_{4}^{+}[z_{2}^{+},z_{4}^{+}]_{\boldsymbol{x}_{2}}^{{\rm cl}}a^{-}(z_{4}^{+},\boldsymbol{x}_{2})[z_{4}^{+},z_{1}^{+}]_{\boldsymbol{x}_{2}}^{{\rm cl}}\label{eq:expanded-line-2}\\
 & -g^{2}\int_{z_{2}^{+}}^{z_{1}^{+}}\!{\rm d}z_{3}^{+}\int_{z_{3}^{+}}^{z_{1}^{+}}\!{\rm d}z_{4}^{+}[z_{2}^{+},z_{3}^{+}]_{\boldsymbol{x}_{2}}^{{\rm cl}}a^{-}(z_{3}^{+},\boldsymbol{x}_{2})[z_{3}^{+},z_{4}^{+}]_{\boldsymbol{x}_{2}}^{{\rm cl}}a^{-}(z_{4}^{+},\boldsymbol{x}_{2})[z_{4}^{+},z_{1}^{+}]_{\boldsymbol{x}_{2}}^{{\rm cl}}\,.\nonumber 
\end{align}
In the r.h.s. of these two relations, the Wilson lines with the superscript
${\rm cl}$ are built solely from classical fields. From now on, we
will omit this superscript and keep in mind that all contributions
from quantum gluons are explicitly written. The quadratic term in
$a$ involved in our dipole operator reads
\begin{align}
 & \frac{1}{g^{2}}{\rm tr}[z_{1}^{+},z_{2}^{+}]_{\boldsymbol{x}_{1}}[z_{2}^{+},z_{1}^{+}]_{\boldsymbol{x}_{2}}\label{eq:expanded-lines-together}\\
 & =\int_{z_{2}^{+}}^{z_{1}^{+}}\!{\rm d}z_{3}^{+}\int_{z_{2}^{+}}^{z_{1}^{+}}\!{\rm d}z_{4}^{+}{\rm tr}[z_{1}^{+},z_{3}^{+}]_{\boldsymbol{x}_{1}}a^{-}(z_{3}^{+},\boldsymbol{x}_{1})[z_{3}^{+},z_{2}^{+}]_{\boldsymbol{x}_{1}}[z_{2}^{+},z_{4}^{+}]_{\boldsymbol{x}_{2}}a^{-}(z_{4}^{+},\boldsymbol{x}_{2})[z_{4}^{+},z_{1}^{+}]_{\boldsymbol{x}_{2}}\nonumber \\
 & -\int_{z_{2}^{+}}^{z_{1}^{+}}\!{\rm d}z_{3}^{+}\int_{z_{3}^{+}}^{z_{1}^{+}}\!{\rm d}z_{4}^{+}{\rm tr}[z_{1}^{+},z_{2}^{+}]_{\boldsymbol{x}_{1}}[z_{2}^{+},z_{3}^{+}]_{\boldsymbol{x}_{2}}a^{-}(z_{3}^{+},\boldsymbol{x}_{2})[z_{3}^{+},z_{4}^{+}]_{\boldsymbol{x}_{2}}a^{-}(z_{4}^{+},\boldsymbol{x}_{2})[z_{4}^{+},z_{1}^{+}]_{\boldsymbol{x}_{2}}\nonumber \\
 & -\int_{z_{2}^{+}}^{z_{1}^{+}}\!{\rm d}z_{3}^{+}\int_{z_{2}^{+}}^{z_{3}^{+}}\!{\rm d}z_{4}^{+}{\rm tr}[z_{1}^{+},z_{3}^{+}]_{\boldsymbol{x}_{1}}a^{-}(z_{3}^{+},\boldsymbol{x}_{1})[z_{3}^{+},z_{4}^{+}]_{\boldsymbol{x}_{1}}a^{-}(z_{4}^{+},\boldsymbol{x}_{1})[z_{4}^{+},z_{2}^{+}]_{\boldsymbol{x}_{1}}[z_{2}^{+},z_{1}^{+}]_{\boldsymbol{x}_{2}}\,.\nonumber 
\end{align}
This way, the difference $\Delta S(x,\boldsymbol{x}_{1},\boldsymbol{x}_{2})$
between the classical part of $S(x,\boldsymbol{x}_{1},\boldsymbol{x}_{2})$
and its one loop correction with 2 quantum gluons reads:
\begin{align}
\Delta S(x,\boldsymbol{x}_{1},\boldsymbol{x}_{2}) & =\frac{g^{2}}{N_{c}}\int{\rm d}z_{1}^{+}\int_{-\infty}^{z_{1}^{+}}\!{\rm d}z_{2}^{+}\,{\rm e}^{ixP^{-}z_{12}^{+}}\frac{\partial^{2}}{\partial z_{1}^{+}\partial z_{2}^{+}}\label{eq:expanded-operator}\\
 & \times\left\{ \int_{z_{2}^{+}}^{z_{1}^{+}}\!{\rm d}z_{3}^{+}\int_{z_{2}^{+}}^{z_{1}^{+}}\!{\rm d}z_{4}^{+}a_{a}^{-}(z_{3}^{+},\boldsymbol{x}_{1})a_{b}^{-}(z_{4}^{+},\boldsymbol{x}_{2}){\rm tr}\left([z_{1}^{+},z_{3}^{+}]_{\boldsymbol{x}_{1}}\boldsymbol{t}^{a}[z_{3}^{+},z_{2}^{+}]_{\boldsymbol{x}_{1}}[z_{2}^{+},z_{4}^{+}]_{\boldsymbol{x}_{2}}\boldsymbol{t}^{b}[z_{4}^{+},z_{1}^{+}]_{\boldsymbol{x}_{2}}\right)\right.\nonumber \\
 & -\int_{z_{2}^{+}}^{z_{1}^{+}}\!{\rm d}z_{3}^{+}\int_{z_{3}^{+}}^{z_{1}^{+}}\!{\rm d}z_{4}^{+}a_{a}^{-}(z_{3}^{+},\boldsymbol{x}_{2})a_{b}^{-}(z_{4}^{+},\boldsymbol{x}_{2}){\rm tr}\left([z_{1}^{+},z_{2}^{+}]_{\boldsymbol{x}_{1}}[z_{2}^{+},z_{3}^{+}]_{\boldsymbol{x}_{2}}\boldsymbol{t}^{a}[z_{3}^{+},z_{4}^{+}]_{\boldsymbol{x}_{2}}\boldsymbol{t}^{b}[z_{4}^{+},z_{1}^{+}]_{\boldsymbol{x}_{2}}\right)\nonumber \\
 & \left.-\int_{z_{2}^{+}}^{z_{1}^{+}}\!{\rm d}z_{3}^{+}\int_{z_{2}^{+}}^{z_{3}^{+}}\!{\rm d}z_{4}^{+}a_{a}^{-}(z_{3}^{+},\boldsymbol{x}_{1})a_{b}^{-}(z_{4}^{+},\boldsymbol{x}_{1}){\rm tr}\left([z_{1}^{+},z_{3}^{+}]_{\boldsymbol{x}_{1}}\boldsymbol{t}^{a}[z_{3}^{+},z_{4}^{+}]_{\boldsymbol{x}_{1}}\boldsymbol{t}^{b}[z_{4}^{+},z_{2}^{+}]_{\boldsymbol{x}_{1}}[z_{2}^{+},z_{1}^{+}]_{\boldsymbol{x}_{2}}\right)\right\} \,.\nonumber 
\end{align}
 The propagator for the quantum gluons in the external classical field
reads:
\begin{equation}
\left\langle a_{a}^{-}(z_{3}^{+},\boldsymbol{x})a_{b}^{-}(z_{4}^{+},\boldsymbol{y})\right\rangle =\frac{1}{4\pi}\int\!\frac{{\rm d}k^{+}}{(k^{+})^{3}}\int\!\frac{{\rm d}^{d}\boldsymbol{k}_{1}}{(2\pi)^{d}}\frac{{\rm d}^{d}\boldsymbol{k}_{2}}{(2\pi)^{d}}{\rm e}^{i(\boldsymbol{k}_{1}\cdot\boldsymbol{x})-i(\boldsymbol{k}_{2}\cdot\boldsymbol{y})}\label{eq:gluon-prop} \left(\boldsymbol{k}_{1}\right|2ik^{+}\delta(z_{34}^{+})\delta^{ab}\boldsymbol{1}+(\boldsymbol{k}_{1}\cdot\boldsymbol{k}_{2}){\cal G}_{k^{+}}^{ab}(z_{3}^{+},z_{4}^{+})\left|\boldsymbol{k}_{2}\right)\,, 
\end{equation}
where the reduced scalar propagator ${\cal G}_{k^{+}}$ satisfies the integrated Schrodinger equations
\begin{equation}
{\cal G}_{k^{+}}^{ab}(z_{3}^{+},z_{4}^{+})={\cal G}_{k^{+}}^{(0)}(z_{3}^{+},z_{4}^{+})\boldsymbol{1}\delta^{ab}+ig\int_{z_{4}^{+}}^{z_{3}^{+}}\!{\rm d}z_{5}^{+}{\cal G}_{k^{+}}^{(0)}(z_{3}^{+},z_{5}^{+})\left[\hat{A}^{-}(z_{5}^{+}){\cal G}_{k^{+}}(z_{5}^{+},z_{4}^{+})\right]^{ab}\,,\label{eq:schroL}
\end{equation}
and
\begin{equation}
{\cal G}_{k^{+}}^{ab}(z_{3}^{+},z_{4}^{+})={\cal G}_{k^{+}}^{(0)}(z_{3}^{+},z_{4}^{+})\boldsymbol{1}\delta^{ab}+ig\int_{z_{4}^{+}}^{z_{3}^{+}}\!{\rm d}z_{5}^{+}\left[{\cal G}_{k^{+}}(z_{3}^{+},z_{5}^{+})\hat{A}^{-}(z_{5}^{+})\right]^{ab}{\cal G}_{k^{+}}^{(0)}(z_{5}^{+},z_{4}^{+})\,.\label{eq:schroR}
\end{equation}
Here and throughout, we will be using Schwinger bracket notations,
see eg~\cite{Blaizot:2015lma} for details. Plugging Eq.~(\ref{eq:gluon-prop})
into Eq.~(\ref{eq:expanded-operator}) yields:
\begin{align}
\Delta S(x,\boldsymbol{x}_{1},\boldsymbol{x}_{2}) & =\frac{\alpha_{s}}{N_{c}}\int{\rm d}z_{1}^{+}\int_{-\infty}^{z_{1}^{+}}\!{\rm d}z_{2}^{+}\,{\rm e}^{ixP^{-}z_{12}^{+}}\frac{\partial^{2}}{\partial z_{1}^{+}\partial z_{2}^{+}}\int\!\frac{{\rm d}k^{+}}{(k^{+})^{3}}\int\!\frac{{\rm d}^{d}\boldsymbol{k}_{1}}{(2\pi)^{d}}\frac{{\rm d}^{d}\boldsymbol{k}_{2}}{(2\pi)^{d}}\label{eq:expanded-operator-with-prop}\\
 & \times\left\{ \int_{z_{2}^{+}}^{z_{1}^{+}}\!{\rm d}z_{3}^{+}\int_{z_{2}^{+}}^{z_{1}^{+}}\!{\rm d}z_{4}^{+}{\rm e}^{i(\boldsymbol{k}_{1}\cdot\boldsymbol{x}_{1})-i(\boldsymbol{k}_{2}\cdot\boldsymbol{x}_{2})}{\rm tr}[z_{1}^{+},z_{3}^{+}]_{\boldsymbol{x}_{1}}\boldsymbol{t}^{a}[z_{3}^{+},z_{2}^{+}]_{\boldsymbol{x}_{1}}[z_{2}^{+},z_{4}^{+}]_{\boldsymbol{x}_{2}}\boldsymbol{t}^{b}[z_{4}^{+},z_{1}^{+}]_{\boldsymbol{x}_{2}}\right.\nonumber \\
 & -\int_{z_{2}^{+}}^{z_{1}^{+}}\!{\rm d}z_{3}^{+}\int_{z_{3}^{+}}^{z_{1}^{+}}\!{\rm d}z_{4}^{+}{\rm e}^{i(\boldsymbol{k}_{1}\cdot\boldsymbol{x}_{2})-i(\boldsymbol{k}_{2}\cdot\boldsymbol{x}_{2})}{\rm tr}[z_{1}^{+},z_{2}^{+}]_{\boldsymbol{x}_{1}}[z_{2}^{+},z_{3}^{+}]_{\boldsymbol{x}_{2}}\boldsymbol{t}^{a}[z_{3}^{+},z_{4}^{+}]_{\boldsymbol{x}_{2}}\boldsymbol{t}^{b}[z_{4}^{+},z_{1}^{+}]_{\boldsymbol{x}_{2}}\nonumber \\
 & \left.-\int_{z_{2}^{+}}^{z_{1}^{+}}\!{\rm d}z_{3}^{+}\int_{z_{2}^{+}}^{z_{3}^{+}}\!{\rm d}z_{4}^{+}{\rm e}^{i(\boldsymbol{k}_{1}\cdot\boldsymbol{x}_{1})-i(\boldsymbol{k}_{2}\cdot\boldsymbol{x}_{1})}{\rm tr}[z_{1}^{+},z_{3}^{+}]_{\boldsymbol{x}_{1}}\boldsymbol{t}^{a}[z_{3}^{+},z_{4}^{+}]_{\boldsymbol{x}_{1}}\boldsymbol{t}^{b}[z_{4}^{+},z_{2}^{+}]_{\boldsymbol{x}_{1}}[z_{2}^{+},z_{1}^{+}]_{\boldsymbol{x}_{2}}\right\} \nonumber \\
 & \times\left(\boldsymbol{k}_{1}\right|2ik^{+}\delta(z_{34}^{+})\delta^{ab}\boldsymbol{1}+(\boldsymbol{k}_{1}\cdot\boldsymbol{k}_{2}){\cal G}_{k^{+}}^{ab}(z_{3}^{+},z_{4}^{+})\left|\boldsymbol{k}_{2}\right)\,.\nonumber 
\end{align}
Note that so-called instantaneous terms with $\delta(z_{34}^{+})$
involve non-logarithmic tadpole integrals, which are set to 0 in dimensional
regularization. We will now detail the computation of the quark self
energy diagram as an example, the others being similar. First, one
needs to rewrite the involved operators thanks to successive uses
of the following relations which are direct consequences of the integrated
Schrodinger equations~(\ref{eq:schroL},\ref{eq:schroR}): for any
$x^{+},y^{+},z^{+},\boldsymbol{x},\boldsymbol{y}$, 
\begin{align}
[x^{+},y^{+}]_{\boldsymbol{y}}(...)[y^{+},x^{+}]_{\boldsymbol{x}} & =(...)+\int_{y^{+}}^{x^{+}}\!{\rm d}z^{+}\partial_{z^{+}}\hat{{\cal I}}_{z^{+}}[z^{+},y^{+}]_{\boldsymbol{y}}(...)[y^{+},z^{+}]_{\boldsymbol{x}}\,,\\
[x^{+},y^{+}]_{\boldsymbol{y}}(...)[y^{+},x^{+}]_{\boldsymbol{x}} & =(...)-\int_{y^{+}}^{x^{+}}\!{\rm d}z^{+}\hat{{\cal I}}_{z^{+}}[x^{+},z^{+}]_{\boldsymbol{y}}(...)[z^{+},x^{+}]_{\boldsymbol{x}}\,,
\end{align}
and
\begin{align}
 {\cal G}_{k^{+}}^{ab}(x^{+},y^{+})[x^{+},y^{+}]_{\boldsymbol{y}}\boldsymbol{t}^{b}[y^{+},x^{+}]_{\boldsymbol{x}} &=\boldsymbol{t}^{a}{\cal G}_{k^{+}}^{(0)}(x^{+},y^{+})+\int_{y^{+}}^{x^{+}}\!{\rm d}z^{+}\hat{{\cal I}}_{z^{+}}{\cal G}_{k^{+}}^{ab}(x^{+},z^{+})[x^{+},z^{+}]_{\boldsymbol{y}}\boldsymbol{t}^{b}[z^{+},x^{+}]_{\boldsymbol{x}}{\cal G}_{k^{+}}^{(0)}(z^{+},y^{+})\,,\\
  [y^{+},x^{+}]_{\boldsymbol{y}}\boldsymbol{t}^{a}[x^{+},y^{+}]_{\boldsymbol{x}}{\cal G}_{k^{+}}^{ab}(x^{+},y^{+})
 &=\boldsymbol{t}^{b}{\cal G}_{k^{+}}^{(0)}(x^{+},y^{+})-\int_{x^{+}}^{y^{+}}\!{\rm d}z^{+}{\cal G}_{k^{+}}^{(0)}(x^{+},z^{+})\hat{{\cal I}}_{z^{+}}{\cal G}_{k^{+}}^{ab}(z^{+},y^{+})[y^{+},z^{+}]_{\boldsymbol{y}}\boldsymbol{t}^{a}[z^{+},y^{+}]_{\boldsymbol{x}}\,,
\end{align}
where we introduced a gluon insertion operator $\hat{{\cal I}}_{z^{+}}$
which for all intents and purposes here, up to subeikonal corrections,
will act as $\partial_{z^{+}}$.

Up to subeikonal corrections, and subtracting the disconnected contributions,
the operator involved in the quark self energy diagrams can be rewritten
as
\begin{align}
 & {\cal G}_{k^{+}}^{ab}(z_{3}^{+},z_{4}^{+}){\rm tr}[z_{1}^{+},z_{3}^{+}]_{\boldsymbol{x}_{1}}\boldsymbol{t}^{a}[z_{3}^{+},z_{4}^{+}]_{\boldsymbol{x}_{1}}\boldsymbol{t}^{b}[z_{4}^{+},z_{2}^{+}]_{\boldsymbol{x}_{1}}[z_{2}^{+},z_{1}^{+}]_{\boldsymbol{x}_{2}}\label{eq:QtoQOperator-1}\\
 & \simeq-C_{F}\left(\int_{z_{2}^{+}}^{z_{4}^{+}}\!{\rm d}z_{5}^{+}\int_{z_{2}^{+}}^{z_{5}^{+}}\!{\rm d}z_{6}^{+}+\int_{z_{3}^{+}}^{z_{1}^{+}}\!{\rm d}z_{5}^{+}\int_{z_{3}^{+}}^{z_{5}^{+}}\!{\rm d}z_{6}^{+}\right){\cal G}_{k^{+}}^{(0)}(z_{3}^{+},z_{4}^{+}){\rm tr}\left\{ \partial_{z_{5}^{+}}\partial_{z_{6}^{+}}[z_{5}^{+},z_{6}^{+}]_{\boldsymbol{x}_{1}}[z_{6}^{+},z_{5}^{+}]_{\boldsymbol{x}_{2}}\right\} \nonumber \\
 & +\int_{z_{4}^{+}}^{z_{3}^{+}}\!{\rm d}z_{5}^{+}\int_{z_{4}^{+}}^{z_{5}^{+}}\!{\rm d}z_{6}^{+}{\cal G}_{k^{+}}^{(0)}(z_{3}^{+},z_{5}^{+}){\rm tr}\left\{ \partial_{z_{5}^{+}}\partial_{z_{6}^{+}}{\cal G}_{k^{+}}^{ab}(z_{5}^{+},z_{6}^{+})\boldsymbol{t}^{a}[z_{5}^{+},z_{6}^{+}]_{\boldsymbol{x}_{1}}\boldsymbol{t}^{b}[z_{6}^{+},z_{5}^{+}]_{\boldsymbol{x}_{2}}\right\} {\cal G}_{k^{+}}^{(0)}(z_{6}^{+},z_{4}^{+})\,,\nonumber 
\end{align}
It is possible to show that the first two terms, which correspond to the virtual corrections in the standard derivations of the BK evolution equation, are subleading in twist, so we can take their fully eikonal limit to reproduce their contribution to BK. This is known, so we will only detail the other term here, denoted with the RBK superscript. We can play the same game with all diagrams, to get:
\begin{align}
\Delta S^{{\rm RBK}}(x,\boldsymbol{x}_{1},\boldsymbol{x}_{2}) & =\frac{\alpha_{s}}{N_{c}}\int{\rm d}z_{1}^{+}\int_{-\infty}^{z_{1}^{+}}{\rm d}z_{2}^{+}{\rm e}^{ixP^{-}z_{12}^{+}}\frac{\partial^{2}}{\partial z_{1}^{+}\partial z_{2}^{+}}\int\!\frac{{\rm d}k^{+}}{k^{+}}\int\!\frac{{\rm d}^{d}\boldsymbol{k}_{1}}{(2\pi)^{d}}\frac{{\rm d}^{d}\boldsymbol{k}_{2}}{(2\pi)^{d}}\frac{\boldsymbol{k}_{1}\cdot\boldsymbol{k}_{2}}{(k^{+})^{2}}\label{eq:}\\
 & \times\int{\rm d}z_{3}^{+}{\rm d}z_{4}^{+}{\rm d}z_{5}^{+}{\rm d}z_{6}^{+}\left\{ \theta(z_{13}^{+})\theta(z_{35}^{+})\theta(z_{56}^{+})\theta(z_{64}^{+})\theta(z_{42}^{+}){\rm e}^{i(\boldsymbol{k}_{1}\cdot\boldsymbol{x}_{1})-i(\boldsymbol{k}_{2}\cdot\boldsymbol{x}_{2})}\right.\nonumber \\
 & \times\left(\boldsymbol{k}_{1}\right|{\cal G}_{k^{+}}^{(0)}(z_{3}^{+},z_{5}^{+}){\rm tr}\left\{ \hat{{\cal I}}_{z_{5}^{+}}\hat{{\cal I}}_{z_{6}^{+}}{\cal G}_{k^{+}}^{ab}(z_{5}^{+},z_{6}^{+})[z_{5}^{+},z_{6}^{+}]_{x_{1}}\boldsymbol{t}^{b}[z_{6}^{+},z_{5}^{+}]_{x_{2}}\boldsymbol{t}^{a}\right\} {\cal G}_{k^{+}}^{(0)}(z_{6}^{+},z_{4}^{+})\left|\boldsymbol{k}_{2}\right)\nonumber \\
 & +\theta(z_{14}^{+})\theta(z_{45}^{+})\theta(z_{56}^{+})\theta(z_{63}^{+})\theta(z_{32}^{+}){\rm e}^{i(\boldsymbol{k}_{1}\cdot\boldsymbol{x}_{1})-i(\boldsymbol{k}_{2}\cdot\boldsymbol{x}_{2})}\nonumber \\
 & \times\left(\boldsymbol{k}_{1}\right|{\cal G}_{k^{+}}^{(0)}(z_{3}^{+},z_{6}^{+}){\rm tr}\left\{ \hat{{\cal I}}_{z_{5}^{+}}\hat{{\cal I}}_{z_{6}^{+}}{\cal G}_{k^{+}}^{ab}(z_{6}^{+},z_{5}^{+})[z_{5}^{+},z_{6}^{+}]_{x_{1}}\boldsymbol{t}^{a}[z_{6}^{+},z_{5}^{+}]_{x_{2}}\boldsymbol{t}^{b}\right\} {\cal G}_{k^{+}}^{(0)}(z_{5}^{+},z_{4}^{+})\left|\boldsymbol{k}_{2}\right)\nonumber \\
 & -\theta(z_{14}^{+})\theta(z_{45}^{+})\theta(z_{56}^{+})\theta(z_{63}^{+})\theta(z_{32}^{+}){\rm e}^{i(\boldsymbol{k}_{1}\cdot\boldsymbol{x}_{2})-i(\boldsymbol{k}_{2}\cdot\boldsymbol{x}_{2})}\nonumber \\
 & \times\left(\boldsymbol{k}_{1}\right|{\cal G}_{k^{+}}^{(0)}(z_{3}^{+},z_{6}^{+}){\rm tr}\left\{ \hat{{\cal I}}_{z_{5}^{+}}\hat{{\cal I}}_{z_{6}^{+}}\boldsymbol{t}^{b}{\cal G}_{k^{+}}^{ab}(z_{6}^{+},z_{5}^{+})[z_{5}^{+},z_{6}^{+}]_{x_{1}}\boldsymbol{t}^{a}[z_{6}^{+},z_{5}^{+}]_{x_{2}}\right\} {\cal G}_{k^{+}}^{(0)}(z_{5}^{+},z_{4}^{+})\left|\boldsymbol{k}_{2}\right)\nonumber \\
 & -\theta(z_{13}^{+})\theta(z_{35}^{+})\theta(z_{56}^{+})\theta(z_{64}^{+})\theta(z_{42}^{+}){\rm e}^{i(\boldsymbol{k}_{1}\cdot\boldsymbol{x}_{1})-i(\boldsymbol{k}_{2}\cdot\boldsymbol{x}_{1})}\nonumber \\
 & \left.\times\left(\boldsymbol{k}_{1}\right|{\cal G}_{k^{+}}^{(0)}(z_{3}^{+},z_{5}^{+}){\rm tr}\hat{{\cal I}}_{z_{5}^{+}}\hat{{\cal I}}_{z_{6}^{+}}\boldsymbol{t}^{a}[z_{5}^{+},z_{6}^{+}]_{x_{1}}\boldsymbol{t}^{b}[z_{6}^{+},z_{5}^{+}]_{x_{2}}{\cal G}_{k^{+}}^{ab}(z_{5}^{+},z_{6}^{+}){\cal G}_{k^{+}}^{(0)}(z_{6}^{+},z_{4}^{+})\left|\boldsymbol{k}_{2}\right)\right\} \,.\nonumber 
\end{align}
The light cone time derivatives are trivial. Taking them and combining terms together yields:
\begin{align}
\Delta S^{{\rm RBK}}(x,\boldsymbol{x}_{1},\boldsymbol{x}_{2}) & =\frac{\alpha_{s}}{N_{c}}\int{\rm d}z_{1}^{+}{\rm d}z_{2}^{+}{\rm e}^{ixP^{-}z_{12}^{+}}\int\!\frac{{\rm d}k^{+}}{k^{+}}\int\!\frac{{\rm d}^{d}\boldsymbol{k}_{1}}{(2\pi)^{d}}\frac{{\rm d}^{d}\boldsymbol{k}_{2}}{(2\pi)^{d}}\frac{\boldsymbol{k}_{1}\cdot\boldsymbol{k}_{2}}{(k^{+})^{2}}\label{eq:-1}\\
 & \times\int{\rm d}z_{5}^{+}{\rm d}z_{6}^{+}\theta(z_{15}^{+})\theta(z_{56}^{+})\theta(z_{62}^{+})\nonumber \\
 & \times\left\{ {\rm e}^{i(\boldsymbol{k}_{1}\cdot\boldsymbol{x}_{1})}\left({\rm e}^{-i(\boldsymbol{k}_{2}\cdot\boldsymbol{x}_{1})}-{\rm e}^{-i(\boldsymbol{k}_{2}\cdot\boldsymbol{x}_{2})}\right)\right.\nonumber \\
 & \times\left(\boldsymbol{k}_{1}\right|{\cal G}_{k^{+}}^{(0)}(z_{1}^{+},z_{5}^{+}){\rm tr}\left\{ \hat{{\cal I}}_{z_{5}^{+}}\hat{{\cal I}}_{z_{6}^{+}}{\cal G}_{k^{+}}^{ab}(z_{5}^{+},z_{6}^{+})[z_{5}^{+},z_{6}^{+}]_{x_{1}}\boldsymbol{t}^{b}[z_{6}^{+},z_{5}^{+}]_{x_{2}}\boldsymbol{t}^{a}\right\} {\cal G}_{k^{+}}^{(0)}(z_{6}^{+},z_{2}^{+})\left|\boldsymbol{k}_{2}\right)\nonumber \\
 & -{\rm e}^{-i(\boldsymbol{k}_{2}\cdot\boldsymbol{x}_{2})}\left({\rm e}^{i(\boldsymbol{k}_{1}\cdot\boldsymbol{x}_{1})}-{\rm e}^{i(\boldsymbol{k}_{1}\cdot\boldsymbol{x}_{2})}\right)\nonumber \\
 & \left.\times\left(\boldsymbol{k}_{1}\right|{\cal G}_{k^{+}}^{(0)}(z_{2}^{+},z_{6}^{+}){\rm tr}\left\{ \hat{{\cal I}}_{z_{5}^{+}}\hat{{\cal I}}_{z_{6}^{+}}{\cal G}_{k^{+}}^{ab}(z_{6}^{+},z_{5}^{+})[z_{5}^{+},z_{6}^{+}]_{x_{1}}\boldsymbol{t}^{a}[z_{6}^{+},z_{5}^{+}]_{x_{2}}\boldsymbol{t}^{b}\right\} {\cal G}_{k^{+}}^{(0)}(z_{5}^{+},z_{1}^{+})\left|\boldsymbol{k}_{2}\right)\right\} \,.\nonumber 
\end{align}
Going back to coordinate space with
\begin{align}
\left(\boldsymbol{k}_{1}\right| & =\int{\rm d}^{d}\boldsymbol{z}_{1}{\rm e}^{-i(\boldsymbol{k}_{1}\cdot\boldsymbol{z}_{1})}\left(\boldsymbol{z}_{1}\right|\label{eq:-2}\\
\left|\boldsymbol{k}_{2}\right) & =\int{\rm d}^{d}\boldsymbol{z}_{2}{\rm e}^{i(\boldsymbol{k}_{2}\cdot\boldsymbol{z}_{2})}\left|\boldsymbol{z}_{2}\right)\nonumber \\
1 & =\int{\rm d}^{d}\boldsymbol{x}_{3}{\rm d}^{d}\boldsymbol{x}_{4}\left|\boldsymbol{x}_{3}\right)\left(\boldsymbol{x}_{3}\right|\left|\boldsymbol{x}_{4}\right)\left(\boldsymbol{x}_{4}\right|\,,\nonumber 
\end{align}
we obtain
\begin{align}
\Delta S^{{\rm RBK}}(x,\boldsymbol{x}_{1},\boldsymbol{x}_{2}) & =\frac{\alpha_{s}}{N_{c}}\int{\rm d}z_{1}^{+}{\rm d}z_{2}^{+}{\rm e}^{ixP^{-}z_{12}^{+}}\int\!\frac{{\rm d}k^{+}}{k^{+}}\int\!\frac{{\rm d}^{d}\boldsymbol{k}_{1}}{(2\pi)^{d}}\frac{{\rm d}^{d}\boldsymbol{k}_{2}}{(2\pi)^{d}}\frac{\boldsymbol{k}_{1}\cdot\boldsymbol{k}_{2}}{(k^{+})^{2}}\int{\rm d}^{d}\boldsymbol{z}_{1}{\rm d}^{d}\boldsymbol{z}_{2}{\rm d}^{d}\boldsymbol{z}_{3}{\rm d}^{d}\boldsymbol{z}_{4}\label{eq:-3}\\
 & \times\int{\rm d}z_{5}^{+}{\rm d}z_{6}^{+}\theta(z_{15}^{+})\theta(z_{56}^{+})\theta(z_{62}^{+}){\rm e}^{-i(\boldsymbol{k}_{1}\cdot\boldsymbol{z}_{1})+i(\boldsymbol{k}_{2}\cdot\boldsymbol{z}_{2})}\nonumber \\
 & \times\left\{ {\rm e}^{i(\boldsymbol{k}_{1}\cdot\boldsymbol{x}_{1})}\left({\rm e}^{-i(\boldsymbol{k}_{2}\cdot\boldsymbol{x}_{1})}-{\rm e}^{-i(\boldsymbol{k}_{2}\cdot\boldsymbol{x}_{2})}\right)\left(\boldsymbol{z}_{1}\right|{\cal G}_{k^{+}}^{(0)}(z_{1}^{+},z_{5}^{+})\left|\boldsymbol{x}_{3}\right)\left(\boldsymbol{x}_{4}\right|{\cal G}_{k^{+}}^{(0)}(z_{6}^{+},z_{2}^{+})\left|\boldsymbol{z}_{2}\right)\right.\nonumber \\
 & \times\left(\boldsymbol{x}_{3}\right|{\rm tr}\left\{ \hat{{\cal I}}_{z_{5}^{+}}\hat{{\cal I}}_{z_{6}^{+}}{\cal G}_{k^{+}}^{ab}(z_{5}^{+},z_{6}^{+})[z_{5}^{+},z_{6}^{+}]_{x_{1}}\boldsymbol{t}^{b}[z_{6}^{+},z_{5}^{+}]_{x_{2}}\boldsymbol{t}^{a}\right\} \left|\boldsymbol{x}_{4}\right)\nonumber \\
 & -{\rm e}^{-i(\boldsymbol{k}_{2}\cdot\boldsymbol{x}_{2})}\left({\rm e}^{i(\boldsymbol{k}_{1}\cdot\boldsymbol{x}_{1})}-{\rm e}^{i(\boldsymbol{k}_{1}\cdot\boldsymbol{x}_{2})}\right)\left(\boldsymbol{z}_{1}\right|{\cal G}_{k^{+}}^{(0)}(z_{2}^{+},z_{6}^{+})\left|\boldsymbol{x}_{3}\right)\left(\boldsymbol{x}_{4}\right|{\cal G}_{k^{+}}^{(0)}(z_{5}^{+},z_{1}^{+})\left|\boldsymbol{z}_{2}\right)\nonumber \\
 & \left.\times\left(\boldsymbol{x}_{3}\right|{\rm tr}\left\{ \hat{{\cal I}}_{z_{5}^{+}}\hat{{\cal I}}_{z_{6}^{+}}{\cal G}_{k^{+}}^{ab}(z_{6}^{+},z_{5}^{+})[z_{5}^{+},z_{6}^{+}]_{x_{1}}\boldsymbol{t}^{a}[z_{6}^{+},z_{5}^{+}]_{x_{2}}\boldsymbol{t}^{b}\right\} \left|\boldsymbol{x}_{4}\right)\right\} \,.\nonumber 
\end{align}
Using the explicit expression for the free propagators
\begin{equation}
\left(\boldsymbol{z}_{i}\right|{\cal G}_{k^{+}}^{(0)}(z_{m}^{+},z_{n}^{+})\left|\boldsymbol{z}_{j}\right)=\int\!\frac{{\rm d}^{d}\boldsymbol{k}}{(2\pi)^{d}}{\rm e}^{i(\boldsymbol{k}\cdot\boldsymbol{z}_{ij})}{\rm e}^{-i\frac{\boldsymbol{k}^{2}-i0}{2k^{+}}z_{mn}^{+}}[\theta(k^{+})\theta(z_{mn}^{+})-\theta(-k^{+})\theta(z_{nm}^{+})]\,,
\end{equation}
we can easily integrate wrt $z_{1}^{+},z_{2}^{+},\boldsymbol{z}_{1},\boldsymbol{z}_{2}$
and transverse momenta, which yields
\begin{align}
\Delta S^{{\rm RBK}}(x,\boldsymbol{x}_{1},\boldsymbol{x}_{2}) & =-\frac{4\alpha_{s}}{N_{c}}\int\!\frac{{\rm d}k^{+}}{k^{+}}\int\!\frac{{\rm d}^{d}\boldsymbol{k}_{1}}{(2\pi)^{d}}\frac{{\rm d}^{d}\boldsymbol{k}_{2}}{(2\pi)^{d}}\int{\rm d}^{d}\boldsymbol{x}_{3}{\rm d}^{d}\boldsymbol{x}_{4}\int{\rm d}z_{5}^{+}\int_{-\infty}^{z_{5}^{+}}\!{\rm d}z_{6}^{+}{\rm e}^{ixP^{-}z_{56}^{+}}\label{eq:-4}\\
 & \times\left\{ \frac{(\boldsymbol{k}_{1}\cdot\boldsymbol{k}_{2}){\rm e}^{i(\boldsymbol{k}_{1}\cdot\boldsymbol{x}_{13})}\left({\rm e}^{-i(\boldsymbol{k}_{2}\cdot\boldsymbol{x}_{14})}-{\rm e}^{-i(\boldsymbol{k}_{2}\cdot\boldsymbol{x}_{24})}\right)}{(\boldsymbol{k}_{1}^{2}-2xP^{-}k^{+}-i0)(\boldsymbol{k}_{2}^{2}-2xP^{-}k^{+}-i0)}\theta(k^{+})\right.\nonumber \\
 & \times\left(\boldsymbol{x}_{3}\right|{\rm tr}\left\{ \hat{{\cal I}}_{z_{5}^{+}}\hat{{\cal I}}_{z_{6}^{+}}{\cal G}_{k^{+}}^{ab}(z_{5}^{+},z_{6}^{+})[z_{5}^{+},z_{6}^{+}]_{x_{1}}\boldsymbol{t}^{b}[z_{6}^{+},z_{5}^{+}]_{x_{2}}\boldsymbol{t}^{a}\right\} \left|\boldsymbol{x}_{4}\right)\nonumber \\
 & -\frac{(\boldsymbol{k}_{1}\cdot\boldsymbol{k}_{2}){\rm e}^{-i(\boldsymbol{k}_{2}\cdot\boldsymbol{x}_{24})}\left({\rm e}^{i(\boldsymbol{k}_{1}\cdot\boldsymbol{x}_{13})}-{\rm e}^{i(\boldsymbol{k}_{1}\cdot\boldsymbol{x}_{23})}\right)}{(\boldsymbol{k}_{1}^{2}+2xP^{-}k^{+}-i0)(\boldsymbol{k}_{2}^{2}+2xP^{-}k^{+}-i0)}\theta(-k^{+})\nonumber \\
 & \left.\times\left(\boldsymbol{x}_{3}\right|{\rm tr}\left\{ \hat{{\cal I}}_{z_{5}^{+}}\hat{{\cal I}}_{z_{6}^{+}}{\cal G}_{k^{+}}^{ab}(z_{6}^{+},z_{5}^{+})[z_{5}^{+},z_{6}^{+}]_{x_{1}}\boldsymbol{t}^{a}[z_{6}^{+},z_{5}^{+}]_{x_{2}}\boldsymbol{t}^{b}\right\} \left|\boldsymbol{x}_{4}\right)\right\}\,. \nonumber 
\end{align}
 In the second term in the brackets, let us do $(k^{+},\boldsymbol{k}_{1},\boldsymbol{k}_{2},\boldsymbol{x}_{3},\boldsymbol{x}_{4})\rightarrow(-k^{+},-\boldsymbol{k}_{2},-\boldsymbol{k}_{1},\boldsymbol{x}_{4},\boldsymbol{x}_{3})$
after noting that
\begin{equation}
\left(\boldsymbol{x}_{3}\right|{\cal G}_{k^{+}}^{ab}(z_{6}^{+},z_{5}^{+})\left|\boldsymbol{x}_{4}\right)=-\left(\boldsymbol{x}_{4}\right|{\cal G}_{-k^{+}}^{ba}(z_{5}^{+},z_{6}^{+})\left|\boldsymbol{x}_{3}\right)\,.\label{eq:-5}
\end{equation}
This allows to recombine all terms with the same operator, so that
\begin{align}
\Delta S^{{\rm RBK}}(x,\boldsymbol{x}_{1},\boldsymbol{x}_{2}) & =-\frac{4\alpha_{s}}{N_{c}}\int\!\frac{{\rm d}k^{+}}{k^{+}}\theta(k^{+})\int\!\frac{{\rm d}^{d}\boldsymbol{k}_{1}}{(2\pi)^{d}}\frac{{\rm d}^{d}\boldsymbol{k}_{2}}{(2\pi)^{d}}\int{\rm d}^{d}\boldsymbol{x}_{3}{\rm d}^{d}\boldsymbol{x}_{4}\int{\rm d}z_{5}^{+}\int_{-\infty}^{z_{5}^{+}}\!{\rm d}z_{6}^{+}{\rm e}^{ixP^{-}z_{56}^{+}}\nonumber \\
 & \times\frac{(\boldsymbol{k}_{1}\cdot\boldsymbol{k}_{2})\left({\rm e}^{i(\boldsymbol{k}_{1}\cdot\boldsymbol{x}_{13})}-{\rm e}^{i(\boldsymbol{k}_{1}\cdot\boldsymbol{x}_{23})}\right)\left({\rm e}^{-i(\boldsymbol{k}_{2}\cdot\boldsymbol{x}_{14})}-{\rm e}^{-i(\boldsymbol{k}_{2}\cdot\boldsymbol{x}_{24})}\right)}{(\boldsymbol{k}_{1}^{2}-2xP^{-}k^{+}-i0)(\boldsymbol{k}_{2}^{2}-2xP^{-}k^{+}-i0)}\label{eq:-6}\\
 & \times\left(\boldsymbol{x}_{3}\right|{\rm tr}\left\{ \hat{{\cal I}}_{z_{5}^{+}}\hat{{\cal I}}_{z_{6}^{+}}{\cal G}_{k^{+}}^{ab}(z_{5}^{+},z_{6}^{+})\boldsymbol{t}^{a}[z_{5}^{+},z_{6}^{+}]_{x_{1}}\boldsymbol{t}^{b}[z_{6}^{+},z_{5}^{+}]_{x_{2}}\right\} \left|\boldsymbol{x}_{4}\right)\,.\nonumber 
\end{align}
The final step here consists of taking the so-called classical approximation:
\begin{align}
\left(\boldsymbol{x}_{3}\right|{\cal G}_{k^{+}}^{ab}(z_{5}^{+},z_{6}^{+})\left|\boldsymbol{x}_{4}\right) & \simeq\left(\boldsymbol{x}_{3}\right|{\cal G}_{k^{+}}^{(0)}(z_{5}^{+},z_{6}^{+})\left|\boldsymbol{x}_{4}\right)[z_{5}^{+},z_{6}^{+}]_{\bar{s}\boldsymbol{x}_{3}+s\boldsymbol{x}_{4}}^{ab}\label{eq:-7}\\
 & =\int\!\frac{{\rm d}^{d}\boldsymbol{k}}{(2\pi)^{d}}{\rm e}^{i(\boldsymbol{k}\cdot\boldsymbol{x}_{34})}{\rm e}^{-i\frac{\boldsymbol{k}^{2}-i0}{2k^{+}}z_{56}^{+}}[z_{5}^{+},z_{6}^{+}]_{\bar{s}\boldsymbol{x}_{3}+s\boldsymbol{x}_{4}}^{ab}\,,\nonumber 
\end{align}
with $s\in[0,1]$ an arbitrary parameter. This allows to integrate
wrt $\boldsymbol{x}_{34}$ and to keep only the dependence on $\boldsymbol{x}\equiv\bar{s}\boldsymbol{x}_{3}+s\boldsymbol{x}_{4}$
\begin{align}
\Delta S^{{\rm RBK}}(x,\boldsymbol{x}_{1},\boldsymbol{x}_{2}) & =-\frac{4\alpha_{s}}{N_{c}}\int\!\frac{{\rm d}k^{+}}{k^{+}}\theta(k^{+})\int\!\frac{{\rm d}^{d}\boldsymbol{k}_{1}}{(2\pi)^{d}}\frac{{\rm d}^{d}\boldsymbol{k}_{2}}{(2\pi)^{d}}\int{\rm d}^{d}\boldsymbol{x}\int{\rm d}z_{5}^{+}\int_{-\infty}^{z_{5}^{+}}\!{\rm d}z_{6}^{+}{\rm e}^{ixP^{-}z_{56}^{+}}\label{eq:-8}\\
 & \times\frac{(\boldsymbol{k}_{1}\cdot\boldsymbol{k}_{2})\left({\rm e}^{i\boldsymbol{k}_{1}\cdot(\boldsymbol{x}_{1}-\boldsymbol{x})}-{\rm e}^{i\boldsymbol{k}_{1}\cdot(\boldsymbol{x}_{2}-\boldsymbol{x})}\right)\left({\rm e}^{-i\boldsymbol{k}_{2}\cdot(\boldsymbol{x}_{1}-\boldsymbol{x})}-{\rm e}^{-i\boldsymbol{k}_{2}\cdot(\boldsymbol{x}_{2}-\boldsymbol{x})}\right)}{(\boldsymbol{k}_{1}^{2}-2xP^{-}k^{+}-i0)(\boldsymbol{k}_{2}^{2}-2xP^{-}k^{+}-i0)}\nonumber \\
 & \times{\rm e}^{-i\frac{(s\boldsymbol{k}_{1}+\bar{s}\boldsymbol{k}_{2})^{2}-i0}{2k^{+}}z_{56}^{+}}{\rm tr}\left\{ \hat{{\cal I}}_{z_{5}^{+}}\hat{{\cal I}}_{z_{6}^{+}}\boldsymbol{t}^{a}[z_{5}^{+},z_{6}^{+}]_{x_{1}}\boldsymbol{t}^{b}[z_{6}^{+},z_{5}^{+}]_{x_{2}}[z_{5}^{+},z_{6}^{+}]_{\boldsymbol{x}}^{ab}\right\} \,.\nonumber 
\end{align}
Renaming dummy variables, approximating $\hat{{\cal I}}$ by $\partial$,
we finally obtain:
\begin{align}
\Delta S^{{\rm RBK}}(x,\boldsymbol{x}_{1},\boldsymbol{x}_{2}) & =-\frac{4\alpha_{s}}{N_{c}}\int\!\frac{{\rm d}k^{+}}{k^{+}}\theta(k^{+})\int\!\frac{{\rm d}^{d}\boldsymbol{k}_{1}}{(2\pi)^{d}}\frac{{\rm d}^{d}\boldsymbol{k}_{2}}{(2\pi)^{d}}\int{\rm d}^{d}\boldsymbol{x}_{3}\nonumber \\
 & \times\frac{(\boldsymbol{k}_{1}\cdot\boldsymbol{k}_{2})\left({\rm e}^{i(\boldsymbol{k}_{1}\cdot\boldsymbol{x}_{13})}-{\rm e}^{i(\boldsymbol{k}_{1}\cdot\boldsymbol{x}_{23})}\right)\left({\rm e}^{-i(\boldsymbol{k}_{2}\cdot\boldsymbol{x}_{13})}-{\rm e}^{-i(\boldsymbol{k}_{2}\cdot\boldsymbol{x}_{23})}\right)}{(\boldsymbol{k}_{1}^{2}-2xP^{-}k^{+}-i0)(\boldsymbol{k}_{2}^{2}-2xP^{-}k^{+}-i0)}\label{eq:RBK-fin}\\
 & \times\int{\rm d}z_{1}^{+}\int_{-\infty}^{z_{1}^{+}}\!{\rm d}z_{2}^{+}{\rm e}^{i\left(xP^{-}-\frac{(s\boldsymbol{k}_{1}+\bar{s}\boldsymbol{k}_{2})^{2}-i0}{2k^{+}}\right)z_{12}^{+}}{\rm tr}\left\{ \partial_{z_{1}^{+}}\partial_{z_{2}^{+}}\boldsymbol{t}^{a}[z_{1}^{+},z_{2}^{+}]_{x_{1}}\boldsymbol{t}^{b}[z_{2}^{+},z_{1}^{+}]_{x_{2}}[z_{1}^{+},z_{2}^{+}]_{\boldsymbol{x}_{3}}^{ab}\right\}\,. \nonumber 
\end{align}
The virtual terms at $x=0$ are known, and read
\begin{align}
\Delta S^{{\rm V}}(x,\boldsymbol{x}_{1},\boldsymbol{x}_{2}) & =-\frac{4\alpha_{s}}{N_{c}}\int\!\frac{{\rm d}k^{+}}{k^{+}}\theta(k^{+})\int\!\frac{{\rm d}^{d}\boldsymbol{k}_{1}}{(2\pi)^{d}}\frac{{\rm d}^{d}\boldsymbol{k}_{2}}{(2\pi)^{d}}\int{\rm d}^{d}\boldsymbol{x}_{3}\label{eq:-9}\\
 & \times\frac{(\boldsymbol{k}_{1}\cdot\boldsymbol{k}_{2})\left({\rm e}^{i(\boldsymbol{k}_{1}\cdot\boldsymbol{x}_{13})}-{\rm e}^{i(\boldsymbol{k}_{1}\cdot\boldsymbol{x}_{23})}\right)\left({\rm e}^{-i(\boldsymbol{k}_{2}\cdot\boldsymbol{x}_{13})}-{\rm e}^{-i(\boldsymbol{k}_{2}\cdot\boldsymbol{x}_{23})}\right)}{\boldsymbol{k}_{1}^{2}\boldsymbol{k}_{2}^{2}}\nonumber \\
 & \times\int{\rm d}z_{1}^{+}\int_{-\infty}^{z_{1}^{+}}\!{\rm d}z_{2}^{+}\left(-2C_{F}\right){\rm tr}\left\{ \partial_{z_{1}^{+}}\partial_{z_{2}^{+}}[z_{1}^{+},z_{2}^{+}]_{x_{1}}[z_{2}^{+},z_{1}^{+}]_{x_{2}}\right\} \,.\nonumber 
\end{align}
Note that they cancel in the leading twist limit. It is thus possible
to modify it by subeikonal corrections with no repercussion on the
collinear part of this evolution equation, so we can add the Gaussian
phase from Eq.~(\ref{eq:RBK-fin}) and the $x$ corrections to the
denominator:
\begin{align}
\Delta S^{{\rm V}}(x,\boldsymbol{x}_{1},\boldsymbol{x}_{2}) & \simeq-\frac{4\alpha_{s}}{N_{c}}\int\!\frac{{\rm d}k^{+}}{k^{+}}\theta(k^{+})\int\!\frac{{\rm d}^{d}\boldsymbol{k}_{1}}{(2\pi)^{d}}\frac{{\rm d}^{d}\boldsymbol{k}_{2}}{(2\pi)^{d}}\int{\rm d}^{d}\boldsymbol{x}_{3}\label{eq:-10}\\
 & \times\frac{(\boldsymbol{k}_{1}\cdot\boldsymbol{k}_{2})\left({\rm e}^{i(\boldsymbol{k}_{1}\cdot\boldsymbol{x}_{13})}-{\rm e}^{i(\boldsymbol{k}_{1}\cdot\boldsymbol{x}_{23})}\right)\left({\rm e}^{-i(\boldsymbol{k}_{2}\cdot\boldsymbol{x}_{13})}-{\rm e}^{-i(\boldsymbol{k}_{2}\cdot\boldsymbol{x}_{23})}\right)}{(\boldsymbol{k}_{1}^{2}-2xP^{-}k^{+}-i0)(\boldsymbol{k}_{2}^{2}-2xP^{-}k^{+}-i0)}\nonumber \\
 & \times\int{\rm d}z_{1}^{+}\int_{-\infty}^{z_{1}^{+}}\!{\rm d}z_{2}^{+}{\rm e}^{i\left(xP^{-}-\frac{(s\boldsymbol{k}_{1}+\bar{s}\boldsymbol{k}_{2})^{2}-i0}{2k^{+}}\right)z_{12}^{+}}C_{F}{\rm tr}\left\{ \partial_{z_{1}^{+}}\partial_{z_{2}^{+}}[z_{1}^{+},z_{2}^{+}]_{x_{1}}[z_{2}^{+},z_{1}^{+}]_{x_{2}}\right\} \,,\nonumber 
\end{align}
Combining real and virtual terms, we can conclude:
\begin{align}
\Delta S(x,\boldsymbol{x}_{1},\boldsymbol{x}_{2}) & =-\frac{4\alpha_{s}}{N_{c}}\int\!\frac{{\rm d}k^{+}}{k^{+}}\theta(k^{+})\int\!\frac{{\rm d}^{d}\boldsymbol{k}_{1}}{(2\pi)^{d}}\frac{{\rm d}^{d}\boldsymbol{k}_{2}}{(2\pi)^{d}}\int{\rm d}^{d}\boldsymbol{x}_{3}\label{eq:-11}\\
 & \times\frac{(\boldsymbol{k}_{1}\cdot\boldsymbol{k}_{2})\left({\rm e}^{i(\boldsymbol{k}_{1}\cdot\boldsymbol{x}_{13})}-{\rm e}^{i(\boldsymbol{k}_{1}\cdot\boldsymbol{x}_{23})}\right)\left({\rm e}^{-i(\boldsymbol{k}_{2}\cdot\boldsymbol{x}_{13})}-{\rm e}^{-i(\boldsymbol{k}_{2}\cdot\boldsymbol{x}_{23})}\right)}{(\boldsymbol{k}_{1}^{2}-2xP^{-}k^{+}-i0)(\boldsymbol{k}_{2}^{2}-2xP^{-}k^{+}-i0)}\nonumber \\
 & \times\int{\rm d}z_{1}^{+}\int_{-\infty}^{z_{1}^{+}}\!{\rm d}z_{2}^{+}{\rm e}^{i\left(xP^{-}-\frac{(s\boldsymbol{k}_{1}+\bar{s}\boldsymbol{k}_{2})^{2}-i0}{2k^{+}}\right)z_{12}^{+}}\nonumber \\
 & \times\left[{\rm tr}\left\{ \partial_{z_{1}^{+}}\partial_{z_{2}^{+}}\boldsymbol{t}^{a}[z_{1}^{+},z_{2}^{+}]_{x_{1}}\boldsymbol{t}^{b}[z_{2}^{+},z_{1}^{+}]_{x_{2}}[z_{1}^{+},z_{2}^{+}]_{\boldsymbol{x}_{3}}^{ab}\right\} -C_{F}{\rm tr}\left\{ \partial_{z_{1}^{+}}\partial_{z_{2}^{+}}[z_{1}^{+},z_{2}^{+}]_{x_{1}}[z_{2}^{+},z_{1}^{+}]_{x_{2}}\right\} \right]\,,\nonumber 
\end{align}
which is our equation \eqn{eq:S2-PTE}.

%%%%%%%%%%%%%%%%%%%%%%%%%%%%%%%%%%%%%%%%%%%%
\section{Transverse integrals \label{app:integrals}}
%%%%%%%%%%%%%%%%%%%%%%%%%%%%%%%%%%%%%%%%%%%%

\subsection{Fourier transform integrals}

The integrals appearing in \secn{sec:he-factorization} are evaluated using the following procedure.
\beq
I_1 &=& \int_{q}  \frac{\bq^i }{\bq^2 }  \left(\rme^{i\bq\cdot\bx_{13}}-\rme^{i\bq\cdot\bx_{23}}\right)=\frac{-i}{(2\pi)^2} \del_{\bx_3}^i\int \frac{\rmd^2 \bq}{\bq^2 }  \left(\rme^{i\bq\cdot\bx_{13}}-\rme^{i\bq\cdot\bx_{23}}\right)\,,\\
&=&  \frac{-i}{2\pi} \del_{\bx_3}^i  \ln\frac{|\bx_{13}|}{|\bx_{23}|} = \frac{i}{2\pi}  \left( \frac{\bx_{13}^i}{\bx_{13}^2}-\frac{\bx_{23}^i}{\bx_{23}^2}\right)\,.
\eeq
and 
\beq
I_2 &=& \int_{k}  \frac{\bk^i }{\bk^2 }  \left(\rme^{-i\bk\cdot\bx_{13}}-\rme^{-i\bk\cdot\bx_{23}}\right)\ln\frac {\bk^2}{2\rho_0^+P^-}= \frac{i}{(2\pi)^2} \del_{\bx_3}^i\int \frac{\rmd^2 \bk}{\bk^2 }  \left(\rme^{-i\bk\cdot\bx_{13}}-\rme^{-i\bk\cdot\bx_{23}}\right)\ln\frac {\bk^2}{2\rho_0^+P^-}\,,\\
&=&  \frac{i}{2\pi)} \del_{\bx_3}^i  \ln\frac{|\bx_{13}|}{|\bx_{23}|} \left( \ln \frac { |\bx_{13}||\bx_{23}|}{2\rho_0^+P^-}  +2 (\gamma_{E}-\ln(2)) \right) \,,\\
&=&  -\frac{i}{(2\pi)}  \left[ \left( \ln \frac { |\bx_{13}||\bx_{23}|}{2\rho_0^+P^-}  +2 (\gamma_{E}-\ln(2)) \right) \left( \frac{\bx_{13}^i}{\bx_{13}^2}-\frac{\bx_{23}^i}{\bx_{23}^2}\right)+\left( \frac{\bx_{13}^i}{\bx_{13}^2}+\frac{\bx_{23}^i}{\bx_{23}^2}\right)  \ln\frac{|\bx_{13}|}{|\bx_{23}|}  \right] \,.
\eeq

\subsection{Proof of equation \eqref{eq:beta=0-x4int}}

To demonstrate the identity Eq.~\eqref{eq:beta=0-x4int}, let us first compute in $d$ dimension the second term, which involves three transverse coordinates $\bx_1,\bx_2,\bx_3$. The trick is to decompose it into a finite term and terms which only depend on two transverse coordinates:
\begin{align}
 \int \rmd^d \bx_4\frac{\bx_{32}^2}{\bx_{34}^2\bx_{24}^2}\ln\left(\frac{\bx_{14}^2}{\bx_{13}^2}\right)&=\int \rmd^d \bx_{43}\left[\frac{2\bx_{42}\cdot\bx_{32}}{\bx_{43}^2\bx_{42}^2}-\frac{1}{\bx_{43}^2}+\frac{1}{\bx_{42}^2}\right]\ln\left(\frac{(\bx_{13}-\bx_{43})^2}{\bx_{13}^2}\right)\,.
\end{align}
Now, the first term is UV and IR finite in $d=2$ and reads
\begin{align}
  \int\rmd^2\bx_{43} \left[\frac{2\bx_{24}\cdot\bx_{23}}{\bx_{43}^2\bx_{42}^2}\right]\ln\left(\frac{(\bx_{13}-\bx_{43})^2}{\bx_{13}^2}\right)&=\pi\ln\left(\frac{\bx_{13}^2}{\bx_{32}^2}\right)\ln\left(\frac{\bx_{13}^2}{\bx_{12}^2}\right)\,,
\end{align}
and the remaining $d$ dimensional integrals can be computed using the standard bubble integral Eq.~\eqref{eq:bubble-int}
\begin{align}
    \int\rmd^d \bx_4 \left[\frac{\bx_{13}^2}{\bx_{14}^2\bx_{34}^2}\ln\left(\frac{\bx_{14}^2}{\bx_{13}^2}\right)+\left(\frac{1}{\bx_{42}^2}-\frac{1}{\bx_{43}^2}\right)\ln\left(\frac{\bx_{14}^2}{\bx_{13}^2}\right)+\frac{\bx_{12}^2}{\bx_{14}^2\bx_{24}^2}\ln\left(\frac{\bx_{13}^2}{\bx_{14}^2}\right)\right]&=-\pi\ln^2\left(\frac{\bx_{12}^2}{\bx_{13}^2}\right)\,.
\end{align}
Combining the two previous equations, we find Eq.~\eqref{eq:beta=0-x4int}. We do not compute the terms involving $\delta$ functions since they do not contribute to the BK equation.
%%%%%%%%%%%%%%%%%%%%%%%%%%%%%%%%%%%%%%%%%%%%%%%%%%%%%%%%%
\section{Derivative of an operator exponential \label{app:Kbar-derivative}}
%%%%%%%%%%%%%%%%%%%%%%%%%%%%%%%%%%%%%%%%%%%%%%%%%%%%%%%%%%%%%%%%%%%%%%

In this appendix, we provide a detailed derivation of \eqn{eq:dexpL}. 
Setting $\abar=1$ to alleviate the notations, we want to compute
\beq\label{eq:dexpL-0}
\frac{\rmd}{\rmd\ln\mu^2}\,\rme^{L(\mu)}\,.
\eeq

It is convenient to express the exponential as a path-ordered exponential,
\begin{align}\label{eq:expL-path}
\rme^{L(\mu)}
&\to \cP \exp\left(\int_0^1 \rmd u\,L(u,\mu)\right)=\sum_{n=0}^{\infty}\!
\int_{0<u_1<\cdots<u_n<1}\!\!\rmd u_1\cdots \rmd u_n\;
L(u_n,\mu)\cdots L(u_1,\mu)\,,
\end{align}
where we keep $L(u,\mu)$ general for the manipulations below and only at the
end specialize to the case $L(u,\mu)=L(\mu)$ (no $u$–dependence).

Differentiating the Dyson series term by term gives
\begin{align}\label{eq:dexpL-series}
&\frac{\rmd}{\rmd\ln\mu^2}\,\cP \exp\left(\int_0^1 \rmd u\,L(u,\mu)\right)\nn
&= \sum_{n=1}^{\infty}\sum_{m=1}^{n}
\int_{0<u_1<\cdots<u_n<1}\,\,\rmd u_1\cdots \rmd u_n\;
L(u_n,\mu)\cdots L(u_{m+1},\mu)\,
\frac{\rmd L(u_m,\mu)}{\rmd\ln\mu^2}\,
L(u_{m-1},\mu)\cdots L(u_1,\mu)\,.
\end{align}

Let $u\equiv u_m$ and set $r\equiv n-m\ge 0$. Using
\beq
\sum_{n=1}^{\infty}\sum_{m=1}^{n}\to \sum_{m=1}^{\infty}\sum_{r=0}^{\infty}\,,
\eeq
we split the ordered product at $u$ into a “left factor”
$u<v_1<\cdots<v_r<1$ and a “right factor” $0<w_1<\cdots<w_{m-1}<u$, and obtain
\begin{align}\label{eq:dexpL-series-2}
\frac{\rmd}{\rmd\ln\mu^2}\,\rme^{L(\mu)}
&= \int_0^1 \rmd u\,\,\left[
\sum_{r=0}^{\infty}\,\,\int_{u<v_1<\cdots<v_r<1}\,\,\rmd v_1\cdots \rmd v_r\;
L(v_r,\mu)\cdots L(v_1,\mu)\right]\,
\frac{\rmd L(u,\mu)}{\rmd\ln\mu^2} \nn
& \times
\left[\sum_{k=0}^{\infty}\,\,\int_{0<w_1<\cdots<w_k<u}\,\,\rmd w_1\cdots \rmd w_k\;
L(w_k,\mu)\cdots L(w_1,\mu)\right]\,.
\end{align}
The two bracketed series are precisely path-ordered exponential on $[u,1]$
and $[0,u]$, respectively, hence
\begin{align}\label{eq:dexpL-result}
\frac{\rmd}{\rmd\ln\mu^2}\,\cP \exp\left(\int_0^1 \rmd u\,L(u,\mu)\right)
= \int_0^1 \rmd u\;
\cP \exp\!\left(\int_u^1 \rmd v\,L(v,\mu)\right)\,
\frac{\rmd L(u,\mu)}{\rmd\ln\mu^2}\,
\cP \exp\left(\int_0^u \rmd w\,L(w,\mu)\right)\,.
\end{align}

If $L$ is independent of $u$, i.e. $L(u,\mu)=L(\mu)$, the path ordering is
trivial and \eqn{eq:dexpL-result} reduces to the standard Duhamel formula
\beq
\frac{\rmd}{\rmd\ln\mu^2}\,\rme^{L(\mu)}
= \int_0^1 \rmd u\;
\rme^{(1-u)L(\mu)}\,
\frac{\rmd L(\mu)}{\rmd\ln\mu^2}\,
\rme^{uL(\mu)}\,.
\eeq

\bibliography{he-references}

\end{document}